\documentclass{article}
\usepackage[a4paper,hmargin=20mm,vmargin=20mm]{geometry}
\usepackage[english]{babel}
\usepackage{siunitx}
\usepackage{xr}  
\usepackage{filecontents}
\makeatletter\@input{xr.tex}\makeatother
\usepackage{amsmath}
\usepackage{amsthm}
\usepackage{amssymb}
\usepackage{mathtools}
\usepackage{dsfont}
\usepackage[normalem]{ulem}
\usepackage{stmaryrd} 

\usepackage[dvipsnames,svgnames,table]{xcolor}

\usepackage{pgfplotstable}
\pgfplotsset{compat=1.18}
\usepackage{csvsimple-legacy}
\csvstyle{mytablestyle}{head to column names,table foot=\bottomrule,}
\usepackage{array, booktabs}
\usepackage{caption}
\usepackage{blkarray} 

\usepackage[titlenumbered,ruled,noend,linesnumbered]{algorithm2e}
\SetKwInOut{Input}{input}
\SetKwInput{Init}{init}
\SetKwFor{FOR}{for}{do}{end For}
\SetKwFor{If}{if}{then}{end}
\SetKwComment{Comment}{//}{}
\SetKw{Return}{return}
\SetNlSty{bfseries}{\color{gray}}{}

\SetKw{Break}{break}

\SetCommentSty{mycommentfont}
\SetEndCharOfAlgoLine{} 

\usepackage{graphicx}
\graphicspath{{./figures/}}
\usepackage[list=true, position=b]{subcaption}
\usepackage[mode=buildnew]{standalone}  
\usepackage{tikz}

\usepackage[authoryear, round]{natbib}

\let\origappendix\appendix 
\renewcommand\appendix{\origappendix}

\definecolor{comment}{HTML}{EB811B}
\newcommand{\tmp}[1]{{#1}_{\mathrm{tmp}}}

\makeatletter
\newcommand*{\rom}[1]{\expandafter\@slowromancap\romannumeral #1@}
\makeatother
\newenvironment{mydescription}[1]
  {\begin{list}{}%
   {\renewcommand\makelabel[1]{##1 \hfill}%
   \settowidth\labelwidth{\makelabel{#1}}%
   \setlength\leftmargin{\labelwidth}
   \addtolength\leftmargin{\labelsep}}}
  {\end{list}}

\definecolor{benoitemagenta}{HTML}{C653A7}
\definecolor{chrom2}{HTML}{EB811B} 
\definecolor{chrom3}{HTML}{3792A4} 

\newcommand{\amelie}[1]{{#1}}

\usepackage{footnotebackref}
\usepackage{hyperref}
\hypersetup{
    pdftitle=sojourntimepdmp,
    colorlinks=true,
    linkcolor=Gray,
    urlcolor=Gray,
    citecolor=Gray
    }

\begin{document}

\begin{center}

{\Large
    {\sc Estimating relapse time distribution from longitudinal biomarker trajectories using iterative regression and continuous time Markov processes}
}
\bigskip

Alice Cleynen$^{1,2}$ \& Benoîte de Saporta$^{1}$ \& Amélie Vernay$^{1}$
\bigskip

{\it
$^{1}$Univ Montpellier, CNRS, Montpellier, France, alice.cleynen@cnrs.fr, benoite.de-saporta@umontpellier.fr, amelie.vernay@umontpellier.fr\\
$^{2}$John Curtin School of Medical Research, The Australian National University, Canberra, ACT, Australia
}
\end{center}

{\bf Abstract.} Biomarker measurements obtained by blood sampling are often used as a non-invasive means of monitoring tumour progression in cancer patients. Diseases evolve dynamically over time, and studying longitudinal observations of specific biomarkers can help to understand patients response to treatment and predict disease progression. We propose a novel iterative regression-based  method to estimate changes in patients status within a cohort that includes censored patients, and illustrate it on clinical data from myeloma cases. We formulate the relapse time estimation problem in the framework of Piecewise Deterministic Markov processes (PDMP), where the Euclidean component is a surrogate biomarker for patient state. This approach enables continuous-time estimation of the status-change dates,  which in turn allows for accurate inference of the relapse time distribution. A key challenge lies in the partial observability of the process, a complexity that has been rarely addressed in previous studies. We evaluate the performance of our procedure through a simulation study and compare it with different approaches. This work is a proof of concept on biomarker trajectories with simple behaviour, but our method can easily be extended to more complex dynamics.

{\bf Keywords.} Piecewise Deterministic Markov Processes, Relapse time, Partial observations, Regression, Censoring

\bigskip
\section{Introduction}
\label{sec:introduction}

\medskip

Biomarker measurements obtained by blood sampling are often used as a non-invasive means of monitoring tumour progression in cancer patients. Diseases evolve dynamically over time, and the study of longitudinal observations of specific biomarkers can help to understand patient response to treatment and predict disease progression. However, to date, most event dates (remission, relapse) are assigned on visit dates and are based on biomarker values exceeding common thresholds or deviating from reference levels, depending on the experience and personal knowledge of each practitioner. It is therefore of interest to develop statistical methods that would automate and refine the assignment of event dates to allow better understanding of the disease progression and improved prognosis of the patients, so that practitioner may adapt decisions. This is a challenging task, as diseases evolve continuously and dynamically over time, whereas biomarker values are only available at discrete times and measured in noise.

Our study focuses on a cohort of patients followed after developing myeloma as part of a study conducted by the Inter-Groupe Francophone du Myélome in 2009 \citep{attal2017lenalidomide}. We aim to estimate their date of entry into remission and date of relapse if any, based on their biomarker levels. The time between these two dates represents the relapse time, the behaviour of which is crucial for understanding disease progression. One of the difficulties of the problem lies in the fact that our observations are noisy and only give partial information: we have a measurement of a biomarker at discrete visit dates, but
its true value at and between dates is unknown.

Longitudinal biomarker observations have been studied in various ways to model disease progression. This includes the early detection of a certain event \citep{amoros_hmm_2019,drescher_screening_2013,han_statistical_2020,tang_biomarker_2017}, or the estimation of a disease phases by detecting discrete changes in disease states \citep{severson_personalized_2020,lorenzi_probabilistic_2019,bartolomeo_progression_2011}, or estimating a consistent ordering of disease events \citep{fonteijn_eventbased_2012}. In \citep{delft_modeling_2022}, the authors compare 9 strategies for the analysis of longitudinal biomarker data.

In most studies focusing on the occurrence of an event of interest, the date of the event is studied alone, in relation to an initial date, often corresponding to the start of follow-up. In our approach, we divide this duration into two phases, namely time to remission relative to the starting point, and time to relapse relative to entry into remission.

\amelie{The simplest method to tackle this problem is thresholding: the time of interest corresponds to the first observation date just before or just after a fixed threshold value is crossed. More sophisticated methods for this include Hidden Markov Models (HMMs) \citep{elliott1997hmm}, change point detection \citep{aminikhanghahi2016survey} or piecewise regression \citep[Section~4.1]{truong2020review}. By default, these methods apply to discrete-time signals and exploit little or no information about the model, which is one of their advantages. However in our context,  we choose a different approach in order to exploit the fact that the underlying biomarker dynamics is time-continuous.}

We propose to embed this duration estimation problem in the framework of Piecewise Deterministic Markov Processes (PDMPs) \citep{davis1984piecewise}. These are continuous-time processes that can handle both continuous \amelie{Euclidean} and discrete \amelie{mode} variables. They are used to describe deterministic motions punctuated by random jumps, and are therefore particularly well suited to our problem. Here we will consider a simple PDMP, and will aim at estimating its hazard rate (jump intensity).

\amelie{Another relevant family of models for continuous-time biological phenomena are switching Ordinary Differential Equations (ODEs) (see for instance \cite{cloez2017probabilistic}). However their dynamics is not as rich as PDMPs. In switching ODEs, the mode is an autonomous Markov chain that determines the evolution of the Euclidean variables, but there are no feedback loops, and the inter-jump times follow exponential distributions that cannot be modulated by the Euclidean variables. In particular, they cannot directly take into account automatic switching when reaching boundaries of the state space or semi-Markov dynamics with interactions between the mode and the Euclidean variables.

}

Statistical estimation for the hazard rate of PDMPs has been studied in the literature, but remains a challenging task in practice. \citet{azais_statistical_2018} gives a general overview of recent methods for statistical parameter estimation in PDMPs based on a wide range of application examples, in both parametric and non-parametric frameworks. Examples of non parametric methods to estimate the jump intensity under different settings and assumptions can be found in, \citet{krell_statistical_2016}, \citet{azais_nonparametric_2014} or \citet{krell_nonparametric_2021} for instance. In all papers, a single trajectory of the process is fully observed in long time. To the best of our knowledge, there exists no method to estimate the hazard rate of a PDMP with hidden jump times\amelie{, discrete noisy observations or censoring (see \citet{azais2025asymptotic} for a recent survey)}.

We will present a new approach to estimate \amelie{remission and relapse dates together with the censorship indicator} of individuals, as well as the distribution of relapse time in a cohort of patients. It is based on two-sided iterative regression and survival analysis.

The remainder of the paper is organized as follows. Section~\ref{sec:data} introduces the data of interest and the statistical problem. In Section~\ref{sec:method}, we define PDMPs and present our proposed framework for relapse time estimation. We conduct a simulation study to evaluate the performance of our method in Section~\ref{sec:simulation}, where we also compare with different existing approaches. The results obtained from applying our estimation method to the real data are presented in Section~\ref{sec:MM}. We conclude with a discussion.

\section{Data description and problem}
\label{sec:data}

\begin{figure}[ht]
\centering
\includegraphics[width=.80\textwidth]{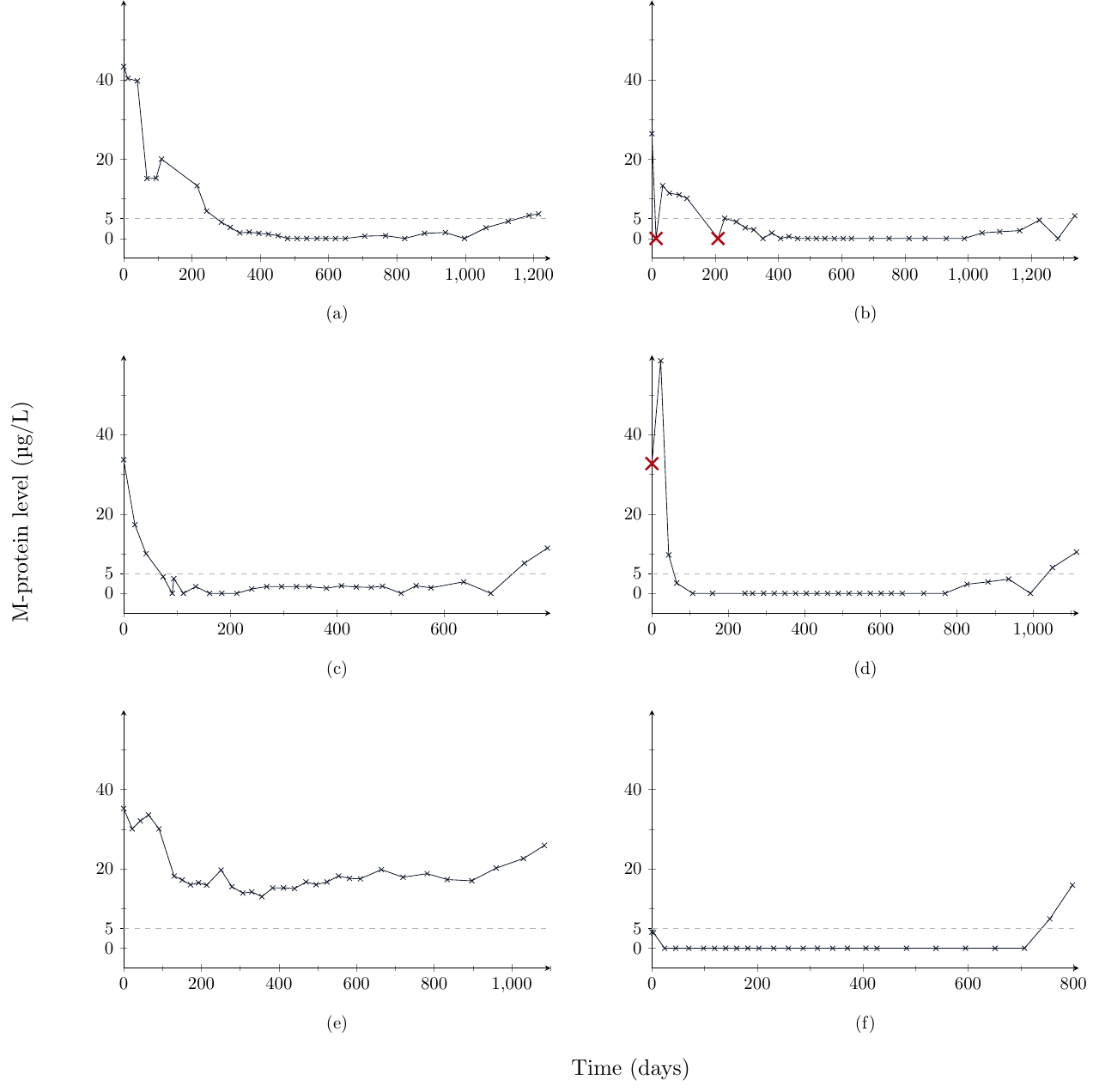}
\caption{\label{fig:trajectories} Some examples of trajectories from the dataset. The crosses correspond to observations and the black lines in between are linear interpolations. Trajectories (a) and (c) are kept as-is. Red crosses in trajectories (b) and (d) corresponds to observations that where removed. Trajectories (e) and (f) are completely removed from the dataset.}
\end{figure}

We are interested in patients undergoing medical follow-up after developing myeloma. Our data come from a \amelie{clinical trial} carried out by the Inter-Groupe Francophone du Myélome in 2009 \citep{attal2017lenalidomide}. A cohort of $763$ patients with newly diagnosed myeloma was followed-up after receiving therapy. Their M-protein levels were measured at different time intervals as a proxy for the progression of their disease --- the higher the level of the biomarker the more severe the condition. At the start of their monitoring, patients have a pathological biomarker level and are administered a treatment, the effect of which is to reduce the M-protein level. If the level falls below a certain fixed threshold, the patient is considered to be in remission. In the event of a relapse, the serum level rises again. Patients may remain in remission, suffer a relapse or leave the study for various reasons. The length of follow-up therefore varies from one individual to another. Biomarker levels are measured during follow-up visits, the frequency of which can vary widely depending on the patient's condition. Some examples of trajectories taken from the dataset are shown in Figure~\ref{fig:trajectories}. Low biomarker levels can be arbitrarily set to zero by practitioners, if considered negligible.

\smallskip

In this work, we are interested in estimating the relapse time of patients, \textit{i.e.} the time elapsed between the date of entry into remission and the date of onset of relapse. It is also called the survival time. The main difficulty of the problem lies in the fact that we only have access to partial observations. The M-protein level evolves continuously but is only measured at discrete dates, and so changes in patient condition are never observed. Added to this is the fact that clinical trial data is always fraught with noise.

\smallskip

The raw data has been preprocessed to remove observations unsuitable for model fitting and the procedure is as follows. We start by iteratively removing the first observation of a trajectory if it is lower than the second one to ensure that the beginning of the trajectory corresponds to a phase when the patient is under treatment (see Fig.~\ref{fig:trajectories}~(d)). We also remove observations with M-protein levels below $1$\si{\micro\gram / \liter} if surrounded by values above $5$\si{\micro\gram / \liter}: we assume that these correspond to measurement errors (see Fig.~\ref{fig:trajectories}~(b)). Then we remove the trajectories whose first observation is lower than a threshold of $5$\si{\micro\gram / \liter}, value at which the level is considered negligible (see Fig.~\ref{fig:trajectories}~(f)) and trajectories with less than two observations below $5$\si{\micro\gram / \liter}: we assume that the associated subjects never really reached remission (see Fig.~\ref{fig:trajectories}~(e)). Finally, we eliminate trajectories with fewer than $10$ observations.

\smallskip

\begin{figure}[ht]
\centering
    \includegraphics[scale=0.30]{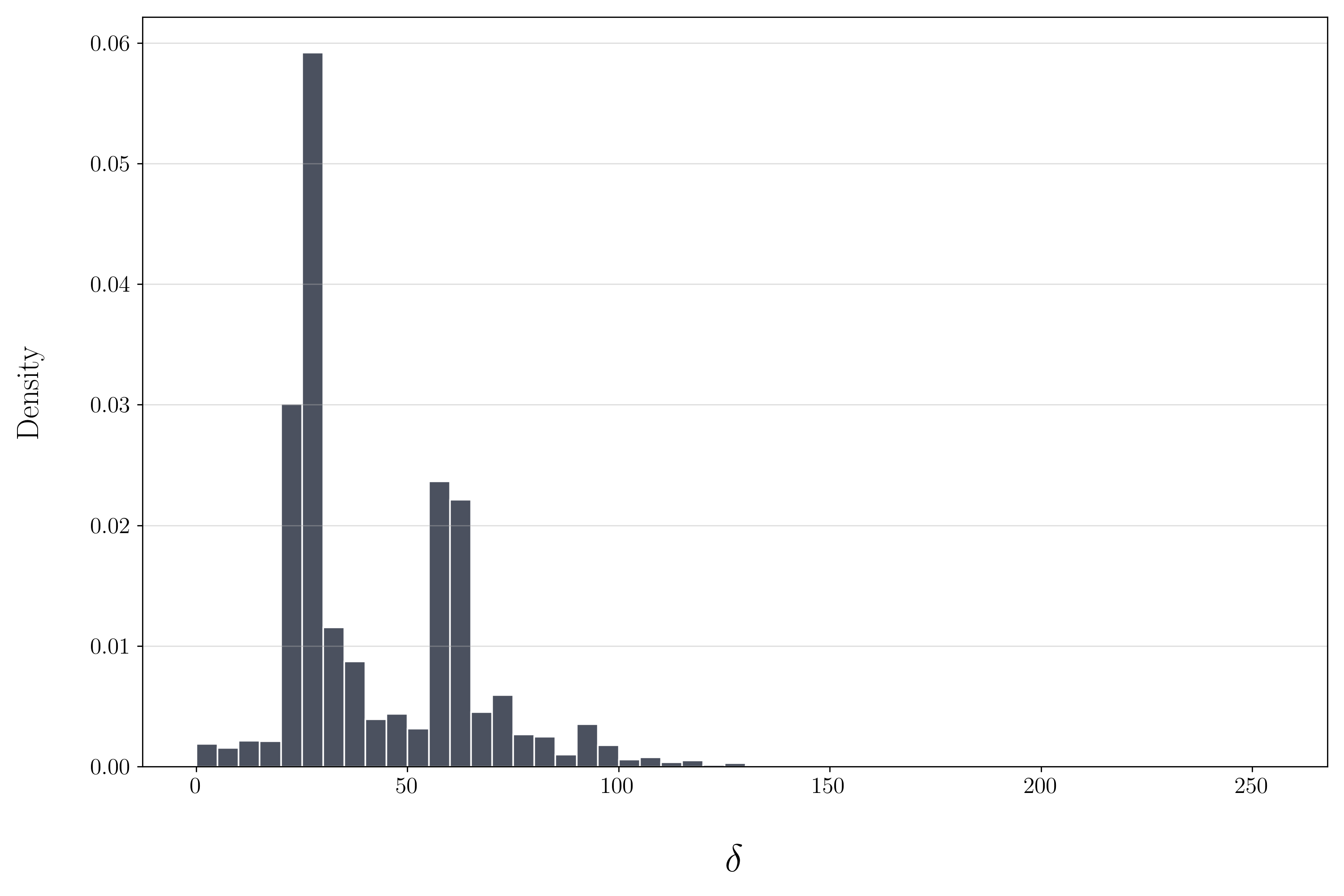}
    \caption{\label{fig:deltas} Distribution of the visit interval $\delta$ (in days) in the application dataset.}
\end{figure}

The post-processed dataset consists of $479$ trajectories, with a mean of about $32$ observations per trajectory. The overall M-protein levels range from $0.0$ to $102.1$. The level at first visit time ranges from $5.01$ to $102.1$ with a median of $34.6$. On average, the last visit occurs after $1406$ days of follow-up. \amelie{In the clinical trial protocol, after an initial premobilization phase, participants are followed for disease progression and survival every month for the first two years following enrollment, then every two months in the absence of adverse events.} 
\amelie{The distribution of visit intervals is represented in Figure~\ref{fig:deltas}. There are two distinct modes, around $\delta=30$ and $\delta=60$ days.}

\section{A new approach based on iterative regression and survival analysis}
\label{sec:method}

\amelie{
In this Section we outline a new approach to estimate relapse time of patients undergoing myeloma treatment from partial and noisy observations. We propose to fit a continuous-time PDMP for the evolution of the biomarker. Our approach contains three main steps:

\begin{enumerate}
    \item We estimate the date of onset of remission and the date of onset of relapse (if any) with iterative curve fitting based on the PDMP. 
    \item Given these dates, we reconstruct the survival duration and the censoring indicator for every trajectory.
    \item We perform a survival analysis to estimate the hazard rate. This can be done with any parametric or non-parametric estimator depending on model assumptions.
\end{enumerate}
}

We have chosen to model our case study using a very simple piecewise-deterministic Markov process, which we describe below. The choice of such a random model offers us simple temporal granularity and interpretable parametrization.

\subsection{PDMPs}
\label{subsec:PDMPs}

Piecewise Deterministic Markov Processes, introduced by Davis in the 80's, are a general class of stochastic processes including almost all non-diffusion Markov models found in applied probability \citep{davis1984piecewise}. These continuous-time processes are used to describe deterministic motions punctuated by random jumps, and are therefore particularly well suited to model our problem. We now give a brief introduction to PDMPs in a generic framework.

\smallskip

\begin{figure}[ht]
\centering
\includegraphics[width=.5\textwidth]{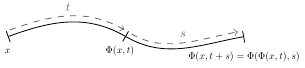}
\caption{\label{fig:flow}Starting from $x$ at time $0$, the process follows its flow up to time $t$ and ends up at $\Phi(x,t)$, assuming no jump has occurred. Going on up to time $t+s$ is the same as starting from $\Phi(x,t)$ and following the flow for a time $s$.}
\end{figure}

Let $(X_t)_{t\geq 0}$ be a PDMP defined on a Borel subset $E\subset \mathbb{R}^d$. The trajectories of $(X_t)_{t\geq 0}$ are determined by the behaviour of the process between jumps, as well as when and where the jumps occur. These aspects are described by a flow $\Phi$, a jump intensity $\lambda$ and a Markov kernel $Q$, respectively. The flow $\Phi \colon E \times \mathbb{R}_{+} \to E$ is a continuous function satisfying the semi-group property: $\forall x\in E,\ \forall t, s\in \mathbb{R}_{+},\ \Phi(x, t+s) = \Phi(\Phi(x,t),s)$. Starting from $x\in E$, $\Phi(x,t)$ gives the position of the process after some time $t$ if no jump has occurred (see Figure~\ref{fig:flow}).

The process can jump deterministically or randomly. Deterministic jumps occur when the flow reaches the boundary $\partial E$ of $E$. Given a starting point $x\in E$, this happens after a time $t^{*}(x) = \inf\left \lbrace t>0\colon \Phi(x,t)\in\partial E \right \rbrace$. Random jumps are governed by the jump intensity $\lambda \colon E \to \mathbb{R}_{+}$ --- also known as the hazard rate --- which is a measurable function such that $\forall x\in E,\ \exists \varepsilon > 0 \colon \int_0^{\varepsilon} \lambda(\Phi(x, s)) \mathrm{d}s < \infty$. That is, jumps cannot occur instantaneously (and therefore there cannot be several jumps at the same time). The jump times of a PDMP are obtained by taking the minimum between deterministic jumps and stochastic ones. Given a starting point $x_0\in E$, for all $t\in \mathbb{R}_{+}$, the first jump time $T_1$ satisfies

\begin{equation} \label{eq:jumptimeproba}
    \mathbb{P}_{X_0=x_0}(T_1 > t) =
    \begin{cases}
     \mathrm{e}^{-\int_0^t \lambda(\Phi(x_0,s))\mathrm{d}s} & \text{if}\quad t < t^{*}(x_0), \\
    0 & \text{if}\quad t\geq t^{*}(x_0).
    \end{cases}
\end{equation}

For both deterministic and random jumps, the new location of the PDMP is drawn from the Markov kernel $Q\colon \bar{E} \times \mathcal{B}(\bar{E}) \to E$, where $\mathcal{B}(\bar{E})$ is the set of Borel $\sigma$-fields of $\bar{E}$, the closure of $E$.
When the process starts from $x\in \bar{E}$, we have that $\forall A\in \mathcal{B}(\bar{E}),\ Q(x, A) = \mathbb{P}(X_{T_1}\in A \mid X_{T_1^{-}} = x)$, where $T_1^{-}$ denotes the time just before the first jump. The Markov kernel $Q$ satisfies $\mathbb{P}(X_{t}=x \mid X_{t^{-}} = x) = 0, \forall t\in\mathbb{R}_{+}$. In other words, each jump must involve a real change of location.

\smallskip

It is common practice to separate the state space $E$ into a hybrid one made up of a discrete component and a continuous one, such that $X_t = (m_t, \zeta_t) \in E \subset M \times \mathbb{R}^d$, where $m_t$ corresponds to a discrete mode and $\zeta_t$ to a continuous variable. Furthermore, the state space can be specific to each mode: $\forall m \in M,\ E_m \subset \mathbb{R}^{d_m}$. The mode-specific flow $\Phi_m$ is such that $\forall m \in M,\ \Phi_m \colon E_m \times \mathbb{R}_{+} \to E_m$ and $\Phi((m,\zeta),t) = (m, \Phi_m(\zeta,t))$. 

\subsection{Model definition and notations}
\label{subsec:model}

\begin{figure}[!ht]
\centering
\includegraphics[width=.7\textwidth]{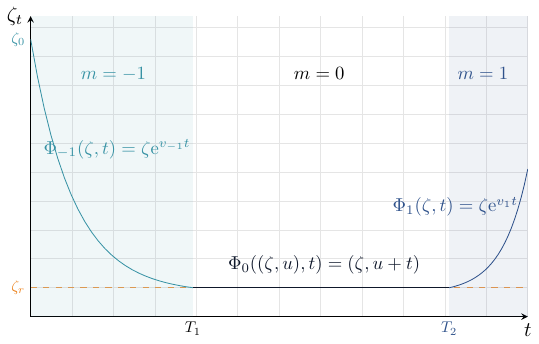}
\caption{\label{fig:pdmp}Example trajectory of our PDMP. The PDMP starts in mode $m=-1$ (patient under treatment) from an initial point $\zeta_0$ and follows a deterministic trajectory along its  exponential flow $\Phi_{-1}$ until the first jump occurs at time $T_1$ when reaching the boundary $\zeta=\zeta_r$ (remission). The mode switches to $m=0$ and the flow is constant equal to $\zeta_r$ until a new jump occurs randomly at time $T_2$. The mode switches then to $m=1$ (relapse of the disease) and the trajectory rises exponentially along the flow $\Phi_1$.}
\end{figure}

The PDMP used to model the dynamics of the biomarker is illustrated in Figure~\ref{fig:pdmp}. At the start of their monitoring patients are administered a treatment, the effect of which is to reduce the M-protein level. If the level falls below a certain fixed threshold $\zeta_r\in\mathbb{R}$, the patient is considered to be in remission. In the event of a relapse, the serum level rises again. The horizon $H$ of follow-up is different for each patient. There are three possible modes for the subjects in the study. They can be sick under treatment ($m=-1$), sick without treatment ($m=1$) or in remission ($m=0$) --- for simplicity, we assume that the process characteristics in remission mode are the same with or without treatment and do not differentiate the two cases. We therefore have $M=\left \lbrace -1, 0, 1 \right \rbrace$ and $E_{-1} = ( \zeta_r, +\infty)$, $E_0 = \left \lbrace \zeta_r \right \rbrace \times \mathbb{R}_{+}$ and $E_{1} = [\zeta_r, +\infty)$. In mode $m=0$, a time variable $u\in\mathbb{R}_{+}$ is added to the state space to allow more flexibility in the jump intensity while ensuring the Markov property holds. It represents the time spent in remission mode. \amelie{Following well accepted disease kinetics in multiple myeloma \cite{sullivan1972kinetics} and other cancers (\cite{blagoev2014therapies})}, under treatment, the biomarker level decreases exponentially with slope $v_{-1}<0$. During relapse, it increases exponentially with slope $v_1>0$. For all $\zeta\in\mathbb{R}$ we have

\begin{equation} \label{eq:modespecificflow} 
\left\{
    \begin{aligned}
        &\Phi_{-1}(\zeta, t) = \zeta\mathrm{e}^{v_{-1}t}, \\
        &\Phi_{0}((\zeta, u), t)  = (\zeta, u+t) = (\zeta_r, u+t) \\
        &\Phi_{1}(\zeta, t)  = \zeta\mathrm{e}^{v_{1}t}.
    \end{aligned}
\right.
\end{equation}

\smallskip

In mode $m=-1$, a jump to $(m=0, \zeta=\zeta_r, u=0)$ occurs when the subject reaches the fixed remission threshold $\zeta_r$. This is therefore a deterministic jump at the boundary. The first jump time $t_{-1}^*(\zeta)$ is the solution of $\Phi_{-1}(\zeta,t)=\zeta_r$. That is, $t_{-1}^*(\zeta) = \frac{1}{v_{-1}}\mathrm{log}(\frac{\zeta_r}{\zeta})$. In mode $m=0$ however, the process may jump randomly to $(m=1, \zeta=\zeta_r)$ and the jump intensity $\lambda_0(u)$ is unknown. \amelie{Note that we make no other specific assumption on the jump intensity.}

We consider that once the process reaches mode $m=1$, no more jump can occur. In our model, the mode-specific Markov kernels are deterministic. When a jump occurs, the biomarker value does not change; only the mode changes, according to a very simple mechanism: at time $t=0$, the process is always in mode $m=-1$ and switches to mode $m=0$ as described above, then possibly to mode $m=1$ if a relapse occurs.

\smallskip

\subsection{Building model estimators}
\label{subsec:estimators}

In this section, we explain the estimation procedure for the parameters of our model based on the data. This involves first estimating the process jump times \amelie{and censoring indicator}, and then finding the parameters of the relapse time distribution. In what follows, we let $t_{-1}^*(\zeta_0)\coloneqq T_1$ be the time of the first jump from mode $m=-1$ to mode $m=0$ and $T_2$ the time of the second jump, if any, from mode $m=0$ to mode $m=1$. Note that even though the underlying process is in continuous time, we only have access to observations at discrete visit dates and the estimation procedure must be adapted accordingly. Note that we use regular time intervals between visits for convenience, but that our method remains valid for any intervals. The points do not need to be equally spaced to perform the estimation, as we will see in Section~\ref{sec:MM}.

\subsubsection{Jump time estimation}
\label{subsubsec:jump time estimation}

Our estimation method for $T_1$ and $T_2$ is an iterative optimisation process based on regression. It is described in Algorithm~\ref{algo:algoparamfit}. Let $(X_t)_{t\geq 0}$ be the PDMP defined in Section \ref{subsec:model} and let $(X_k)_{k\in\mathbb{N}}=(X_{d_k})_{k\in\mathbb{N}}$ be the process at the observation dates $(d_k)_{k\in\mathbb{N}}$. The observations are defined as $Y_k=F(X_k)\mathrm{e}^{\varepsilon_k}$, where $\varepsilon_k \sim \mathcal{N}(0, \sigma^2)$ is a Gaussian noise and where $F\colon E\to\mathbb{R}_{+}$ is a function that returns the marker component of the PDMP. Hence, $Y_k = \zeta_k \mathrm{e}^{\varepsilon_k}$. We use a multiplicative noise both to match the exponential growth and decay of biomarker level and to simplify the estimation procedure described hereafter. We start with the estimation of $T_1$.

\medskip

\textbf{Fitting the flow to observations}\quad For a given $3\leq j \leq N $, let $y_{1:j}=(y_1, y_2,\dots, y_j)$ be the first $j$ values of an $N$-length trajectory of M-protein levels recorded at dates $d_1, d_2,\dots d_j$, respectively. The biomarker level has an exponential form, so we use least squares to fit a linear function $f\colon x\mapsto ax+b$ to the logarithm of our data, and we let $\widetilde{a}=\widetilde{v_{-1}}$ and $\mathrm{e}^{\widetilde{b}}=\widetilde{\zeta_0}$ be the optimal solutions of the problem. We use a logarithmic transformation to prevent errors at the beginning of the trajectory from having too much weight on the overall error.

\medskip

\textbf{Computing the fit error}\quad We then estimate $T_1$ as the solution for $t$ of $\mathrm{e}^{\widetilde{b}}\mathrm{e}^{\widetilde{a}t} = \zeta_r$ (see Alg.~\ref{algo:algoparamfit} l.~\ref{algoline:findT1}). This gives us an approximation $\widetilde{T_1}$ of the jump time from $m=-1$ to $m=0$ and we calculate a general regression error $\Delta_{\mathrm{tmp}}$ as the sum of two errors: one between the points falling before $\widetilde{T_1}$ and the fitted curve, and the other on the remaining part of the trajectory. That is,

\begin{equation}
    \Delta_{\mathrm{tmp}} = \lVert y_{1:k} - \widetilde{b}\mathrm{e}^{-d_{1:k} \times \widetilde{a}} \rVert_2^2 + \lVert y_{k+1:N} - \zeta_r \rVert_2^2,
\end{equation}

where $k = \lvert\{ i ; d_i \leq \widetilde{T_1} \}\rvert$ represents the number of points falling before $\widetilde{T_{1}}$.

\medskip

\textbf{Repeating until convergence}\quad Note that all the above estimates depend on $j$, which is omitted for clarity. This process is repeated for $j\in\left \lbrace 3,\dots,N \right \rbrace$ until a stopping criterion is met, minimizing the error. This results in estimates $\widehat{T_1}$ and $\widehat{a}$ for the entry time into remission $T_1$ and the slope $v_{-1}$ in mode $m=-1$, respectively. Note that $\widehat{a}$ is only used to estimate $T_1$ and will not be used to estimate the relapse time afterwards. Details of the estimation procedure can be found in Algorithm~\ref{algo:algoparamfit}. The last condition on line \ref{algoline:condition} of the algorithm ends the estimation process after $15$ visits if the fitting error stops decreasing to avoid unnecessary iterations.

\medskip

\textbf{Estimating the second jump time}\quad We can then use the same process again on the remaining part of the trajectory --- that is, on $(y_k)_{\widehat{T_1}< k\leq N}$ --- to obtain $\widehat{T_2}$ and $\widehat{v_1}$. \\

\textbf{Censoring criterion}\quad A subject is declared censored (no relapse detected) if either of the following conditions holds on the post-remission trajectory: (i) the 5 most recent M-protein measurements of the post-remission follow-up are all below $3\,\si{\micro\gram\per\litre}$, indicating that the patient appears to remain in remission at the end of their follow-up; or (ii) the estimated relapse slope satisfies $\widehat{v_1} < v_{\min}$, where $v_{\min} = 0.005$ is a minimum detectable slope threshold. Note that condition~(i) operates on the reversed post-remission trajectory, so it checks whether the 5 observations \emph{closest to the end of follow-up} are all below $3\,\si{\micro\gram\per\litre}$ --- this is distinct from the remission threshold $\zeta_r = 1\,\si{\micro\gram\per\litre}$, and reflects that a patient whose most recent biomarker values show no sign of elevation is likely still in remission. Below $v_{\min}$, the trajectory is indistinguishable from a flat signal and no reliable relapse onset can be identified. These two conditions together constitute the sole model selection step in our procedure. In such cases, we assume that no change in mode occurred; the subject is considered censored and their survival time is taken as the time from $\widehat{T_1}$ to the end of follow-up. This criterion is discussed further in Section~\ref{sec:comparisons}.

\begin{algorithm}[ht]
    \caption{Jump time estimation}\label{algo:algoparamfit}
    \Input{$d\in\mathbb{R}^N$ vector of visit dates, $y\in\mathbb{R}^N$ vector of observations at visit dates, $\zeta_r\in\mathbb{R}$ theoretical threshold for remission mode}

    \Init{$\Delta = \infty$, $\widehat{T_1}=0$, $\widehat{a}=0$, $\widehat{b}=0$}
    \BlankLine
    \FOR{$j=3,\dots,N$}{
        $\tmp{d} = d_{1:j}$  \Comment*{slicing of the first $j$ coordinates of $d$}
        $\tmp{y} = y_{1:j}$ \\
        find $\widetilde{a}$ and $\widetilde{b}$ optimal values when fitting a linearization of $f:x\mapsto a\mathrm{e}^{-bx}$ to $\mathrm{log}(\tmp{y})$ using least squares \\
        $\widetilde{T_{1}} = (\mathrm{log}(\zeta_r) - \mathrm{log}(\widetilde{b})) / \widetilde{a}$ \Comment*{solve $\widetilde{b}\mathrm{e}^{t\widetilde{a}} = \zeta_r$ for $t$}\label{algoline:findT1}
        $k = \lvert \{i ; d_i \leq \widetilde{T_1}\} \rvert$ \\
        $n_1 = \lVert y_{1:k} - \widetilde{b}\mathrm{e}^{-d_{1:k} \times \widetilde{a}} \rVert_2^2$ \Comment*{error between the first $k$ points and the fitted curve}
        $n_2 = \lVert y_{k+1:N} - \zeta_r \rVert_2^2$ \Comment*{error on the remaining part of the trajectory}
        $\tmp{\Delta} = n_1 + n_2$ \\
        \If{$\tmp{\Delta} \leq \Delta$ $\mathrm{AND}$ $\widetilde{T_1} > 0$}{
            $\Delta = \tmp{\Delta}$ \\
            $\widehat{T_1} = \widetilde{T_1}$ \\
            $\widehat{a} = \widetilde{a}$ \\
            $\widehat{b} = \widetilde{b}$ \\
        }
        \lIf{$\tmp{\Delta} > \Delta$ $\mathrm{AND}$ $\widetilde{T_1} > 0$ $\mathrm{AND}$ $j > 15$}{break}

    }\label{algoline:condition}
    \Return{$\widehat{T_1}, \widehat{a}, \widehat{b}$}
\end{algorithm}

\subsubsection{Survival time before relapse and censoring indicator}
\label{subsubsec:Survival time before relapse}

Having calculated $\widehat{T_1}$ and possibly $\widehat{T_2}$ for each trajectory, we now have access to the survival times of the subjects,  that is, the elapsed time between the start of the remission and the beginning of the relapse, if any, or the follow-up time otherwise. Following the terms of survival analysis, an \emph{event} is defined as the occurrence of a relapse. A patient is considered \emph{censored} if no event has occurred until the end of its follow-up. \amelie{Our method also gives access to the censoring indicator. For a trajectory in which no $T_2$ is detected, we consider the patient censored.} The survival function $S(t) = \mathbb{P}(T_2-T_1 > t)$ \amelie{$=\mathrm{e}^{-\int_{0}^{t} \lambda_0(s) \mathrm{d}s}$} gives us the probability that a patient remains in remission beyond a time $t$ after remission entry. \amelie{Our aim is to estimate this survival function.}

\medskip 

\amelie{Our estimation method is essentially piecewise regression, but instead of estimating jump times jointly, we first estimate $T_1$, then $T_2$ together with the censoring indicator. We found that this approach performs better in the presence of censoring, see Section~\ref{sec:simulation} for details.}

\section{Simulation study}
\label{sec:simulation}

We evaluate the performance of our estimation strategy through a simulation study and explore the influence of nuisance parameters on a range of scenarios. We then compare our approach with existing methods for estimating relapse time.

The average computation time for the full estimation process (two-sided trajectory fitting and survival regression) for $500$ samples is \SI{7.2}{\second}. All codes are executed on a laptop computer with an Intel(R) Core(TM) i$7$-$12700$H processor using \SI{16}{\giga\byte} of RAM.

\subsection{Generating data}

We use Algorithm \ref{algo:algosimulation} to simulate trajectories. It simulates a trajectory according to the PDMP model described in Section~\ref{subsec:model}, and adds additive noise. The threshold $\zeta_r$ for remission mode, the first M-protein level $\zeta_0$, the time of last observation $H$, as well as the slopes $v_{-1}$ and $v_1$ are all chosen to produce trajectories that are qualitatively similar to those observed in the application dataset presented in Section \ref{sec:data}. 

\amelie{For this specific simulation study, we chose to use a Weibull distribution for the jump intensity. The probability density function of the Weibull distribution is given by

\begin{equation} \label{eq:weibullpdf}
    f(u) = \left( \frac{\alpha}{\beta} \right) \left( \frac{u}{\beta} \right)^{\alpha-1} \mathrm{e}^{-\left( \frac{u}{\beta} \right)^{\alpha}},
\end{equation}

for $u>0$ and where $\alpha>0$ is a shape parameter and $\beta>0$ is a scale parameter. The intensity is therefore the hazard function

\begin{equation} \label{eq:weibullhazard}
    \lambda_0(u) = \left( \frac{\alpha}{\beta} \right) \left( \frac{u}{\beta} \right)^{\alpha-1}.
\end{equation}

Note that a shape parameter $\alpha < 1$ (resp. $\alpha > 1$) means that the failure rate decreases (resp. increases) over time. If $\alpha = 1$, this rate is constant. Again, parameters $\alpha$ and $\beta$ are chosen to produce trajectories that are similar to those of the myeloma dataset.
}

The nuisance parameters, \textit{i.e.} the time interval $\delta$ between visits and the level of noise $\sigma$, together with the number $n$ of trajectories in a cohort are studied over \amelie{four} different scenarios.

\begin{mydescription}{\bfseries SCENARIO}
    \setlength\itemsep{1em}
    
    \item[\textbf{Scenario~\rom{1}}] The number $n$ of trajectories and the noise level $\sigma$ are fixed while the number of days between visit varies, with $\delta \in \left\lbrace 10, 20, 30, 40, 50, 60 \right\rbrace$.
    
    \item[\textbf{Scenario~\rom{2}}] The number of trajectories $n$ and the visit interval $\delta$ are fixed while the noise level varies, with $\sigma \in \left\lbrace 0.25, 1, 2.5, 5 \right\rbrace$.

    \item[\textbf{Scenario~\rom{3}}] The noise level $\sigma$ and the visit interval $\delta$ are fixed while the number of trajectories varies, with $n \in \left\lbrace 100, 500, 1000, 5000, 10000 \right\rbrace$.

    \amelie{\item[\textbf{Scenario~\rom{4}}] The number $n$ of trajectories and the noise level $\sigma$ are fixed while the number of days between visits and the slope parameters are drawn from discrete and continuous uniform distributions, respectively.}
\end{mydescription}

The parameters used in the simulations for each scenario are summarized in Table \ref{table:scenarios}.

\begin{table}[ht]
\caption{\label{table:scenarios} Parameters used to simulate trajectories under the \amelie{four} scenarios.}
\centering
\begin{tabular}{lcccc}
\toprule
\multicolumn{1}{c}{} & \multicolumn{1}{c}{\textbf{Scenario~\rom{1}}} & \multicolumn{1}{c}{\textbf{Scenario~\rom{2}}} & \multicolumn{1}{c}{\textbf{Scenario~\rom{3}}} & \multicolumn{1}{c}{\textbf{\amelie{Scenario~\rom{4}}}} \\
\cmidrule(rl){2-2} \cmidrule(rl){3-3} \cmidrule(rl){4-4} \cmidrule(rl){5-5}
$n$               &                                               $500$ &                                        $500$ & $\left\lbrace 250, 500, 1000, 5000, 10000 \right\rbrace$ & \amelie{$500$} \\
$\delta$          & $\left\lbrace 10, 20, 30, 40, 50, 60 \right\rbrace$ &                                         $30$ & $30$ & \amelie{$\mathcal{U}_{\llbracket 10,60 \rrbracket}$} \\
$\sigma$          &                                                 $1$ & $\left\lbrace 0.25, 1, 2.5, 5 \right\rbrace$ & $1$ & \amelie{$1$} \\
$\zeta_0$         &                 $\mathcal{U}_{\left[15, 55\right]}$ &          $\mathcal{U}_{\left[15, 55\right]}$ & $\mathcal{U}_{\left[15, 55\right]}$ & \amelie{$\mathcal{U}_{\left[15, 55\right]}$} \\
$H$               &              $\mathcal{U}_{\left[900, 1900\right]}$ &       $\mathcal{U}_{\left[900, 1900\right]}$ & $\mathcal{U}_{\left[900, 1900\right]}$ & \amelie{$\mathcal{U}_{\left[900, 1900\right]}$} \\
$(\alpha, \beta)$ &                                      $(4.69, 1650)$ &                               $(4.69, 1650)$ & $(4.69, 1650)$ & \amelie{$(4.69, 1650)$} \\
$v_{-1}$          &                                            $-0.046$ &                                     $-0.046$ & $-0.046$ & \amelie{$\mathcal{U}_{\left[-0.057, -0.028\right]}$} \\
$\zeta_{r}$       &                                                 $1$ &                                          $1$ & $1$ & \amelie{$1$} \\
$v_{1}$           &                                             $0.012$ &                                      $0.012$ & $0.012$ & \amelie{$\mathcal{U}_{\left[0.008, 0.019\right]}$} \\
\bottomrule
\end{tabular}
\end{table}

\begin{algorithm}[ht]
    \caption{Simulation of one trajectory from a PDMP}
    \label{algo:algosimulation}
    \Input{
        $l_0, u_0 \in\mathbb{R}$, $l_H, u_H \in\mathbb{R}$, lower and upper bounds for starting point and follow-up time distribution, $\alpha, \beta\in\mathbb{R}$ shape and scale parameters for the Weibull distribution, $v_{-1}, v_1\in\mathbb{R}$ slopes for mode $m=-1$ and $m=1$, $\delta\in\mathbb{N}$ number of days between two visit dates, $\sigma\in\mathbb{R}^{+}$ standard deviation of the Gaussian noise, $\zeta_r\in\mathbb{R}$ theoretical threshold for remission mode
    }

    \Init{
        $\zeta_{0} \sim \mathcal{U}_{\left[l_0, u_0\right]}$, $H \sim \mathcal{U}_{\left[l_H, u_H\right]}$
    }
    
    \BlankLine

    $T_1 \leftarrow (\mathrm{log}(\zeta_r) - \mathrm{log}(\zeta_{0})) / v_{-1}$ ; $w \sim \mathcal{W}(\alpha, \beta)$ \\
    $T_2 \leftarrow T_1 + w$ ; $c \leftarrow \mathds{1}_{\left \lbrace H \leq T_2 \right \rbrace}$ \Comment*{c censoring indicator}
    $\delta_{\mathrm{end}} \leftarrow \lfloor H/\delta \rfloor$ \Comment*{last visit date}
    \FOR{$k=0,\dots,\delta_{\mathrm{end}}$}{
        $d_k \leftarrow k\delta$ \Comment*{visit dates at regular time intervals until $H$}
        $m_k \leftarrow -\mathds{1}_{\left \lbrace d_k < T_1 \right \rbrace} + \mathds{1}_{\left \lbrace d_k > T_2 \right \rbrace}$ \\
    
        \uIf{$m_k = -1$}{
            $\zeta_k \leftarrow \Phi_{-1}(\zeta_r, d_k)$ \;
        }
        \uElseIf{$m_k = 0$}{
          $\zeta_k \leftarrow \zeta_r$ \;
        }
        \Else{
          $\zeta_k \leftarrow \Phi_{1}(\zeta_r, d_k-T_2)$ \;
        }
        $y_k = \zeta_k + \mathcal{N}(0, \sigma^2)$
    }

    \BlankLine

    \Return{$y = (y_k)_{k}$}
\end{algorithm}

\amelie{Note that in the simulation study, the censoring is dictated by the initial marker value (which determines deterministically $T_1$), the choice of the survival distribution (which determines $T_2=T_1+W$), and the horizon choice (which indicates whether $T_2$ is smaller than $H$).} 
\amelie{A batch of trajectories contains on average $66.7\%$ of censored cases, which again is close to the rate estimated on real data.}

\subsection{Evaluation of the method}

We present here the most relevant results. The online supplementary material contains all the results for the four scenarios. For each experiment, our method is evaluated on 100 batches of $n$ trajectories to take account of variability.

We evaluate our method by examining the errors made on the estimated parameters, namely $T_1$, $T_2$, $\alpha$ and $\beta$, as well as on the censoring prediction. Jump time estimates are compared with the true parameters in absolute distance. Weibull parameter estimates are compared with the true parameters in relative distance, since the two are of different orders of magnitude. We do not take into account the distance between distributions, as we use a parametric estimation method.

\medskip

\textbf{Scenario~\rom{1}} (visit intervals)

\begin{figure}[ht]
\centering
\captionbox{\label{fig:jumpscenario1} \textbf{Scenario~I.} Boxplots of signed estimation errors for $\widehat{T_1}$ and $\widehat{T_2}$ (one batch, $n=500$) as a function of visit interval $\delta$ (days). Errors are computed only on trajectories where a true relapse was correctly detected. Positive values indicate the event was estimated later than its true time.}
[.45\textwidth]{\includegraphics[scale=0.20]{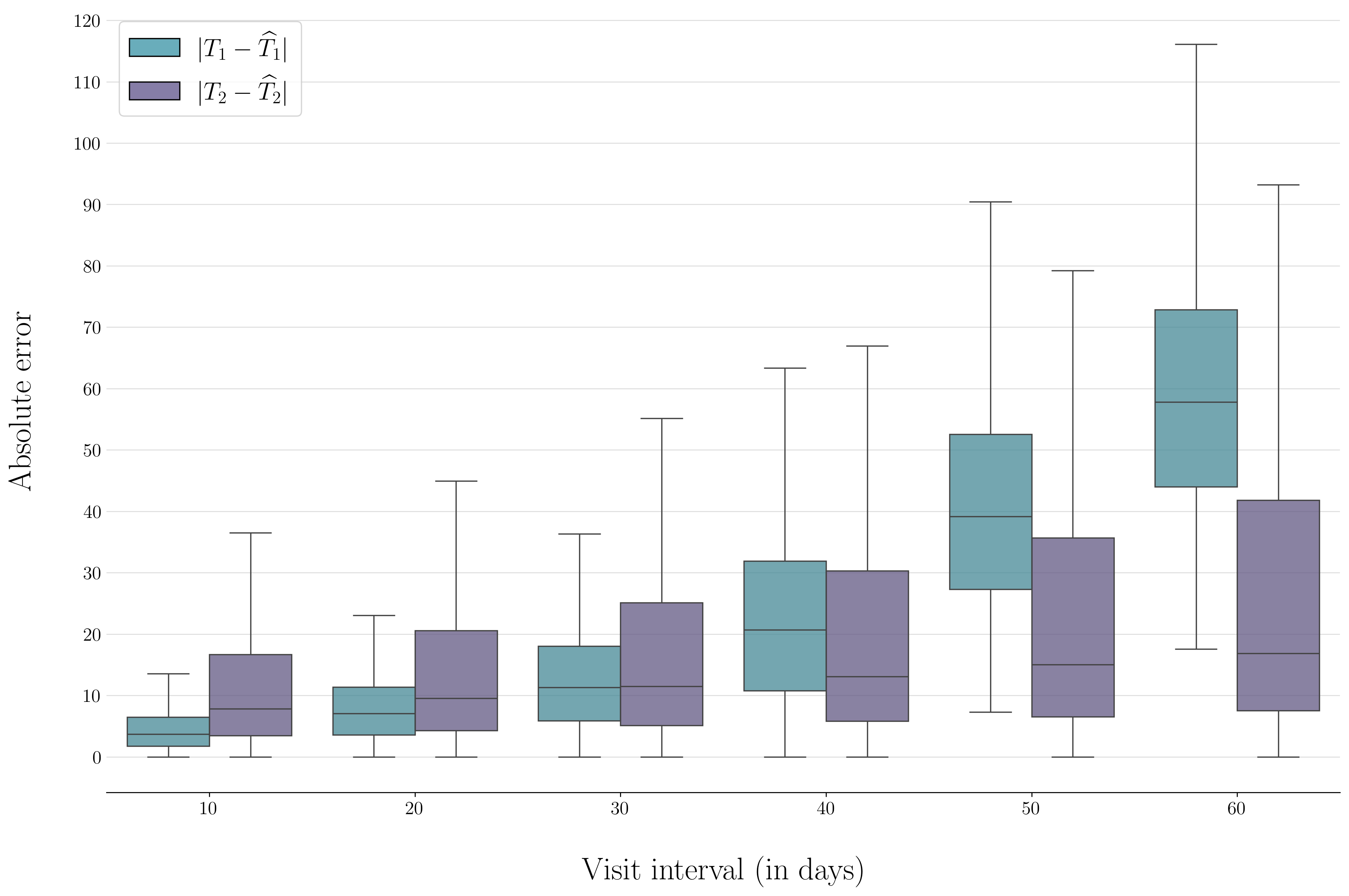}}\hspace{1em}%
\captionbox{\label{fig:weibullparamscenario1} \textbf{Scenario~I.} Relative errors on Weibull shape ($\alpha$) and scale ($\beta$) parameter estimates across $100$ batches ($n=500$) as a function of visit interval $\delta$ (days). Errors are normalised by the true parameter value.}
[.45\textwidth]{\includegraphics[scale=0.20]{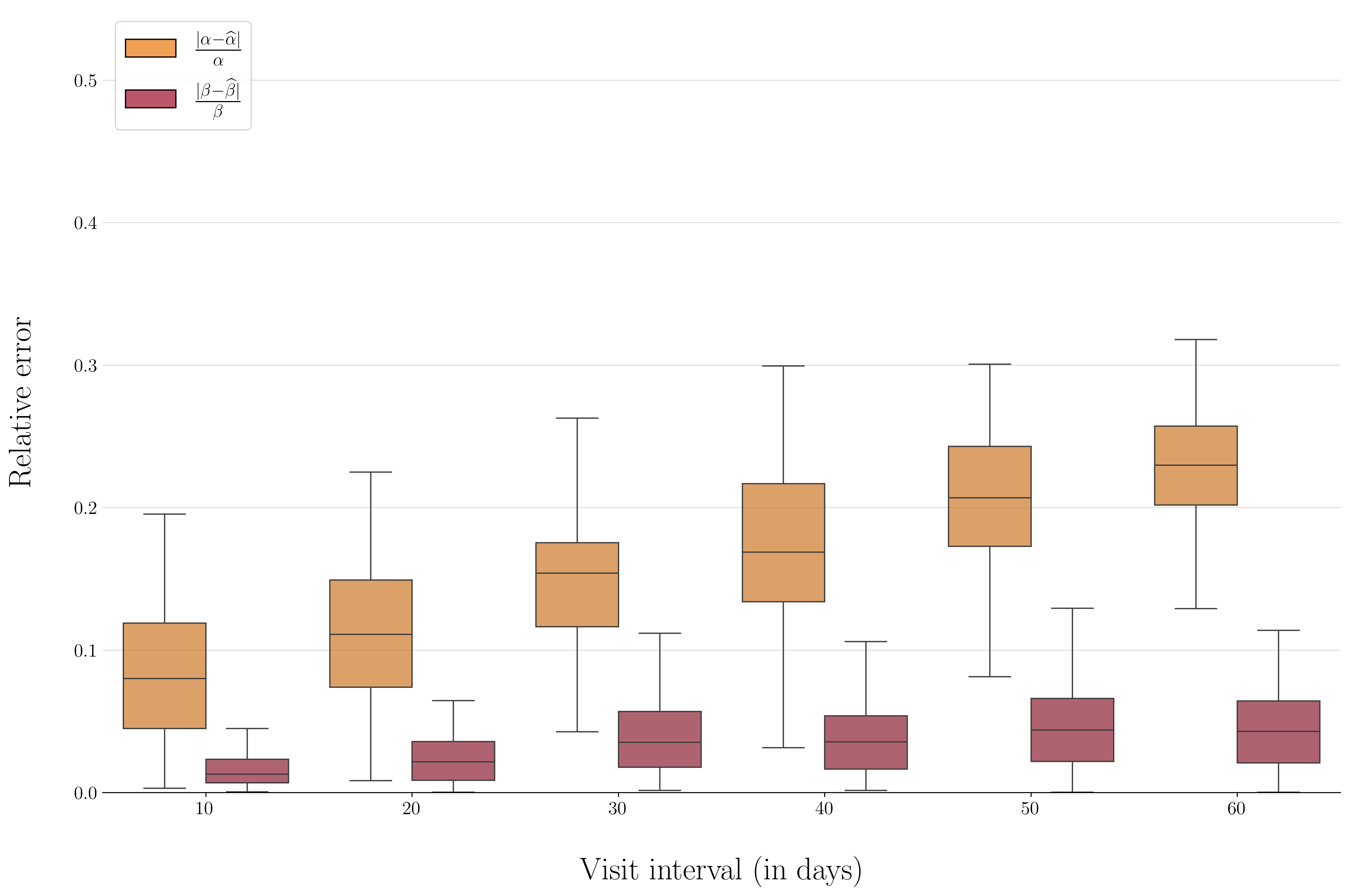}}
\end{figure}

Figure~\ref{fig:jumpscenario1} presents the distributions of absolute errors on the estimates of $T_1$ and $T_2$ depending on visit frequency. Unsurprisingly, with longer time intervals between visits both the mean error and the variability on $\widehat{T_1}$ increase, as fewer points are available to fit the trajectory. For $\widehat{T_2}$ on the other hand, the evolution of error and variability with $\delta$ is less obvious. Note that the errors shown in the figure only concern trajectories for which a relapse has occurred and been correctly detected. We can thus assume that these are more obvious relapses, and therefore relapses for which $T_2$ is easier to estimate. For $\delta=50$ and $\delta=60$, the average number of days of error on the first jump time is close to the time interval itself: the estimate $\widehat{T_1}$ is on average one visit apart from the actual jump time $T_1$. For $\delta \leq 40$ however, the error on $\widehat{T_1}$ is about half the value of the time interval. For $\delta \leq 20$, the mean error on $\widehat{T_1}$ is less that the one on $\widehat{T_2}$, whereas the opposite occurs for larger values of $\delta$. Absolute errors on overall relapse times $T_2 - T_1$ are available in Table~\ref{supp-tab:survivalerrorscenario1} of the supplementary materials. Average survival time for patients who relapse is around $1100$ days, hence errors on these durations are relatively small. \amelie{Table~\ref{supp-tab:relapseexceedsdelta} of the supplementary materials shows the percentage of times the error on survival time exceeded the visit interval $\delta$. Even in the worst case, the error is lower than the visit interval more than $80\%$ of the time.}

\begin{figure}[ht]
\centering
\captionbox{\label{fig:histfreq30} \textbf{Scenario~I, $\delta=30$ days.} Empirical distribution of estimated relapse times $\widehat{T_2}-\widehat{T_1}$ for one batch ($n=500$). The fitted Weibull density (solid) and ground-truth Weibull density (dashed) are overlaid. Only correctly detected relapses are included.}
[.45\textwidth]{\includegraphics[width=0.30\textwidth]{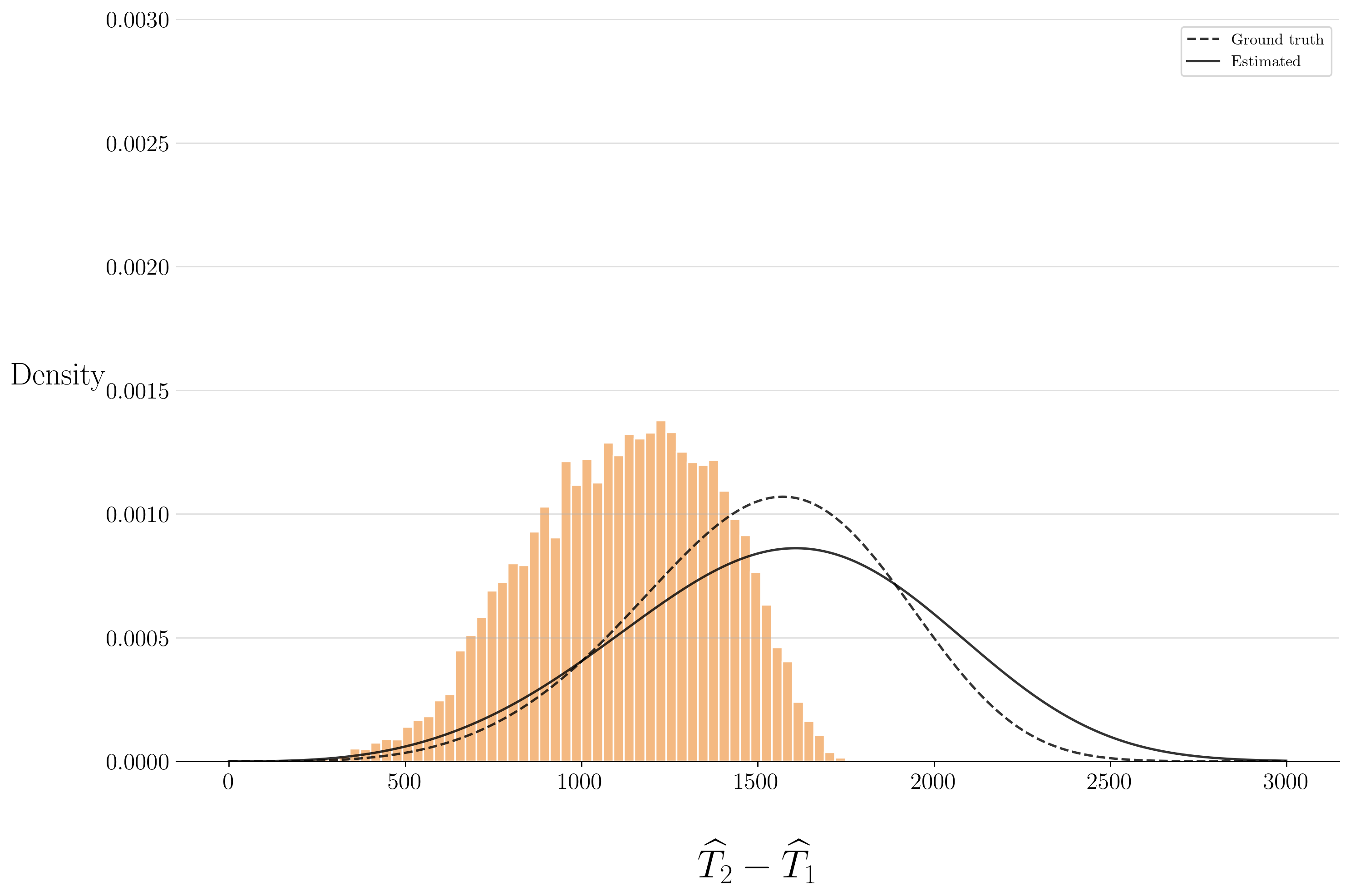}}\hspace{1em}%
\captionbox{\label{fig:ctfreq30} \textbf{Scenario~I, $\delta=30$ days.} Average confusion matrix over $100$ batches ($n=500$). Row rates are percentages of actual relapsed (top) and actual censored (bottom) subjects. Values are mean $\pm$ standard deviation across batches.}
[.45\textwidth]{\includegraphics[width=0.30\textwidth]{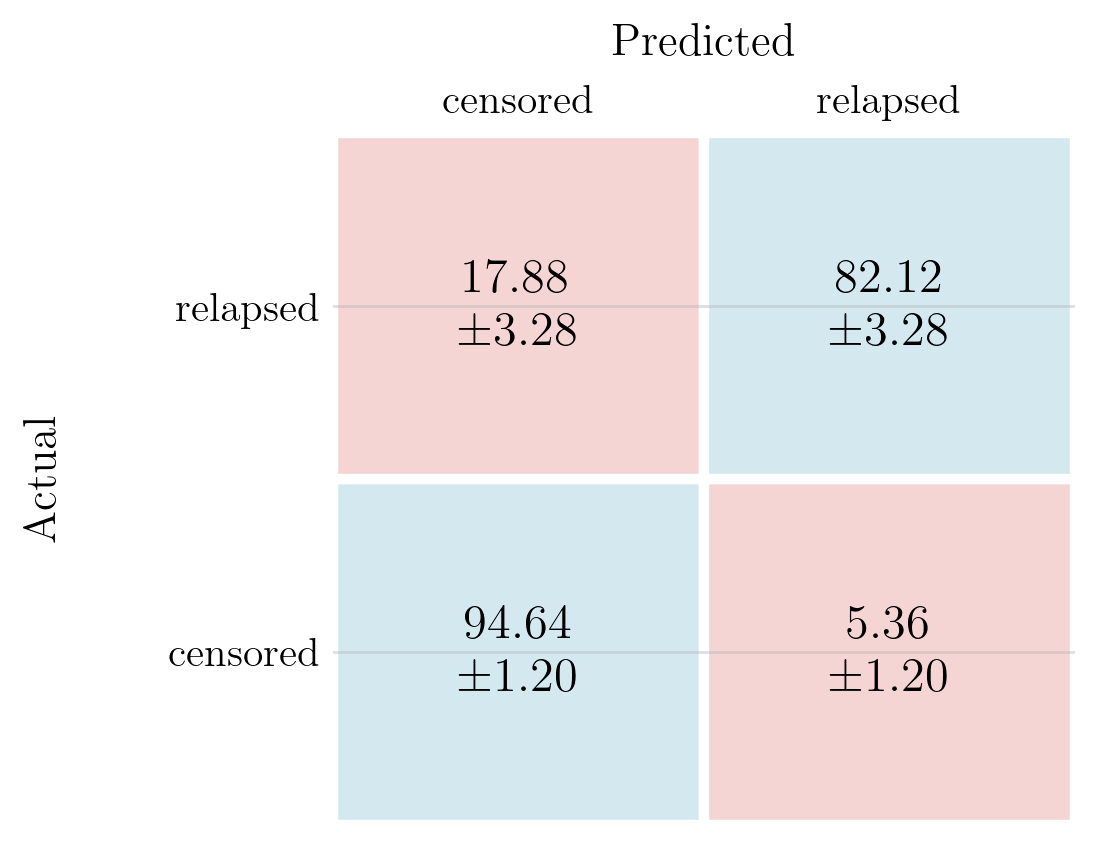}}
\end{figure}

Figure~\ref{fig:weibullparamscenario1} shows the distributions of relative errors on Weibull shape and scale parameter estimates. The average error on the shape parameter $\alpha$ increases with the time between visits. This is fairly consistent with the results in Figure~\ref{fig:jumpscenario1}, since increasing $\delta$ degrades the estimate of both jump times and therefore spreads out the distribution of relapse time. The average relative error on $\widehat{\beta}$ remains constant as the time between visits increases. Figure~\ref{fig:histfreq30} shows the empirical distribution of relapse times for $\delta=30$, together with the Weibull probability density function with shape and scale parameters fitted on the trajectories and the ground-truth density. The estimated distribution has a slightly lower mode than the true one and is more spread out. \amelie{Note, however, that with an average censored case rate of $66.7\%$, our method is inevitably biased.} This aspect will be developed further below. The same curves for other values of $\delta$ are shown in Figure~\ref{supp-fig:pdfsscenario1} of the supplementary materials. Figure~\ref{fig:ctfreq30} represents the average confusion matrix from the estimation on $100$ batches of trajectories with $\delta=30$. A false censoring occurs when the $T_2$ estimates falls after the time horizon $H$. Such errors tend to appear more often as the time interval between visits increases (see Figure~\ref{supp-fig:confusiontablesscenario1} in supplementary materials): the longer we wait before checking a patient again, the more likely we are to miss a relapse. Our method maintains a low false relapse prediction rate: with the censoring threshold, approximately $5\%$ of truly censored subjects are incorrectly classified as relapsed at $\delta=30$ days.

\medskip

\textbf{Scenario~\rom{2}} (noise level)

\begin{figure}[ht]
\centering
\captionbox{\label{fig:jumpscenario2} \textbf{Scenario~II.} Signed estimation errors for $\widehat{T_1}$ and $\widehat{T_2}$ (one batch, $n=500$) as a function of noise level $\sigma$. Errors computed only on correctly detected relapses.}
[.45\textwidth]{\includegraphics[scale=0.20]{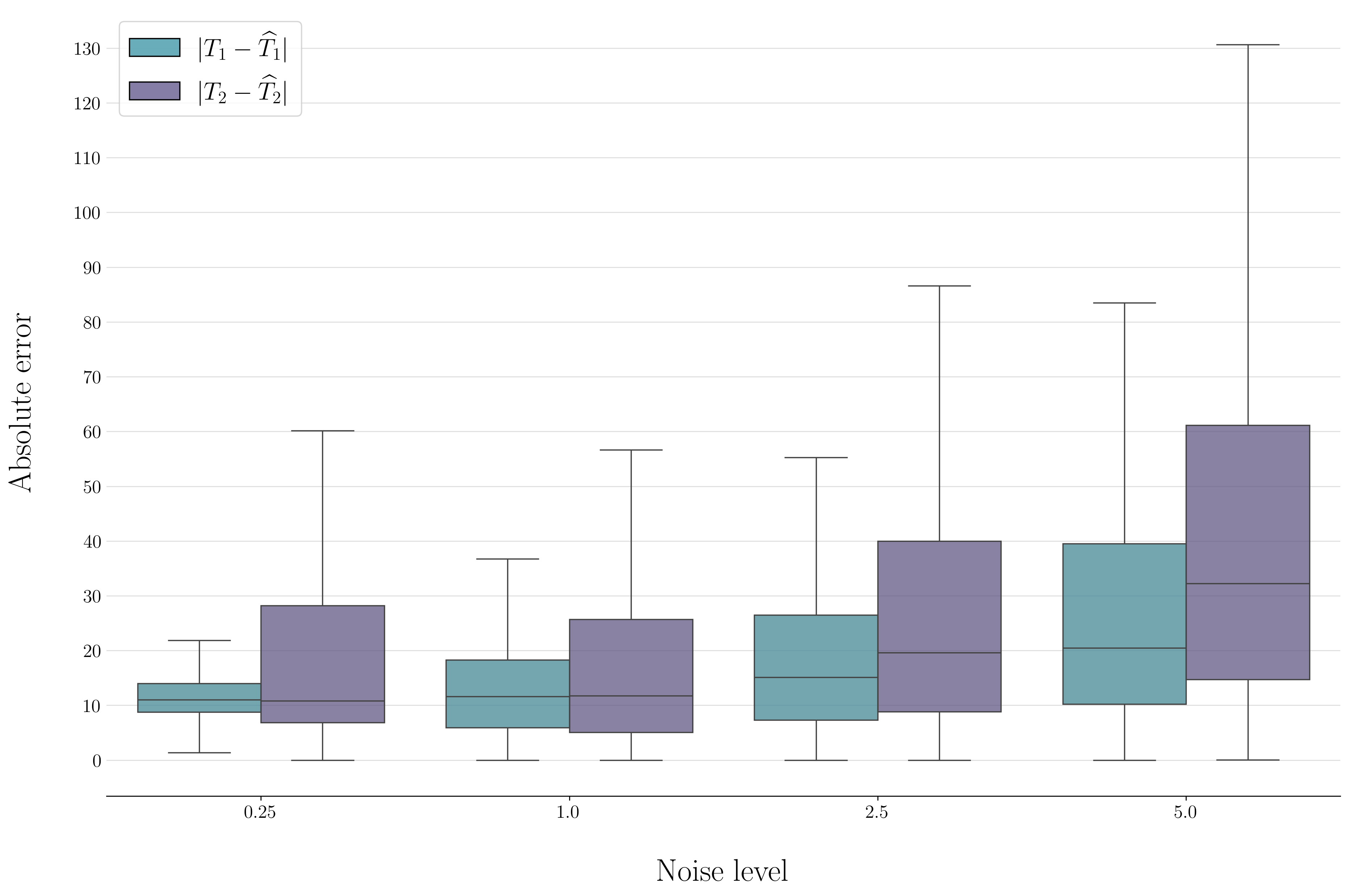}}\hspace{1em}%
\captionbox{\label{fig:weibullparamscenario2} \textbf{Scenario~II.} Relative errors on Weibull shape ($\alpha$) and scale ($\beta$) parameter estimates across $100$ batches ($n=500$) as a function of noise level $\sigma$.}
[.45\textwidth]{\includegraphics[scale=0.20]{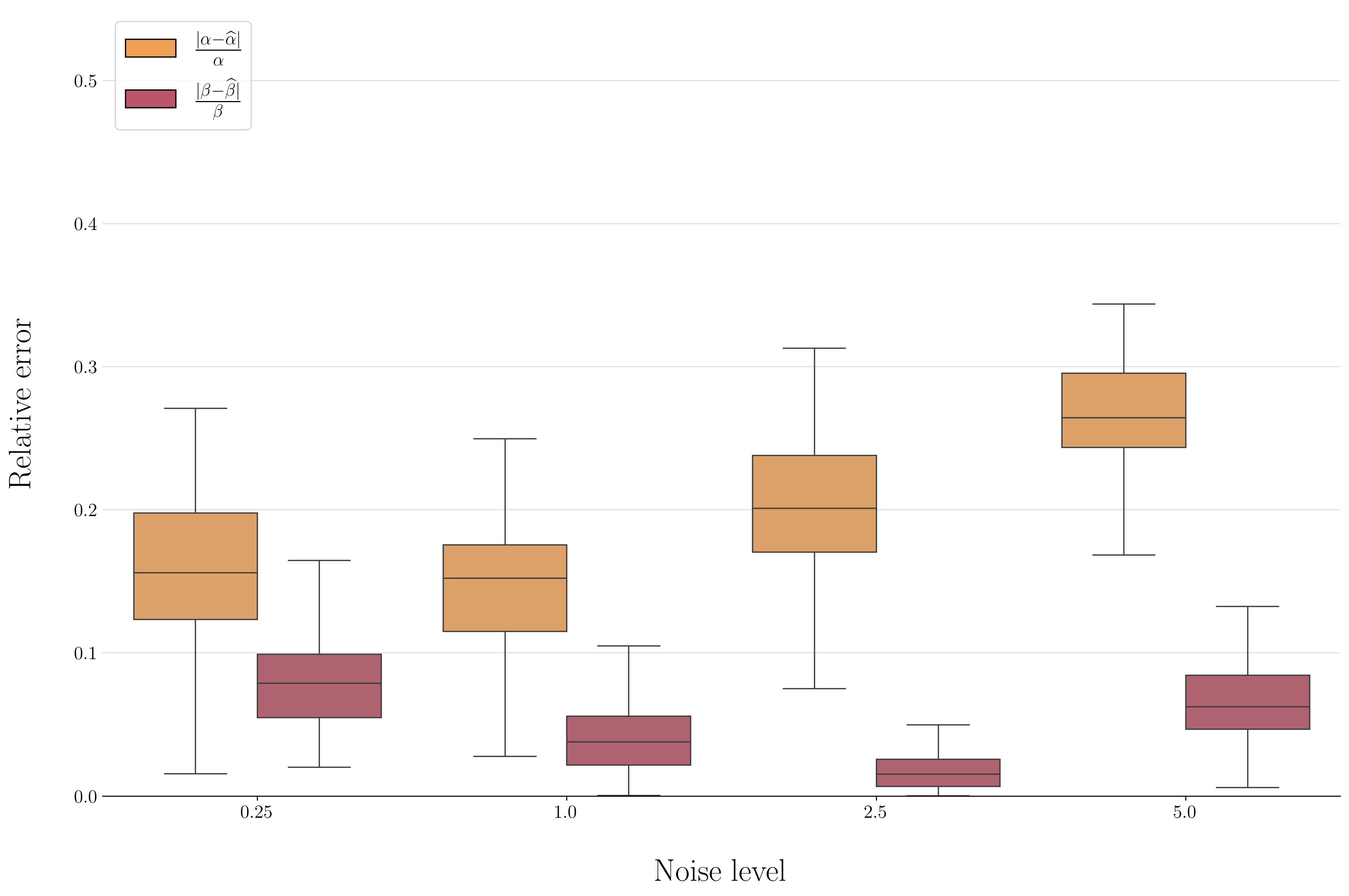}}
\end{figure}

Table~\ref{tab:confusiontablesscenario2} shows the sensitivity and specificity for different noise levels. The noisier the trajectory, the more difficult it is to detect true relapses. This is also illustrated in Figure~\ref{fig:jumpscenario2}, where we see a deterioration in the $T_2$ estimate as $\sigma$ increases. Figure~\ref{fig:weibullparamscenario2} suggests that the estimation errors on jump times have a direct impact on the estimation of the shape parameter $\alpha$ of the relapse time distribution. On the other hand, these errors do not seem to affect the estimation of $\beta$. Overall, the increase in noise level only slightly alters the estimates. Note that for $\sigma=2.5$ and $\sigma=5$, the noise level is much higher than that likely to be found in the trajectories of the application dataset. This scenario shows that the limitation of our estimation method does not stems from the amount of noise in the data.

\begin{table}[ht]
\centering
\caption{\label{tab:confusiontablesscenario2} \textbf{Scenario~II.} Sensitivity and specificity of PDMP over $100$ batches, depending on noise level $\sigma$ ($n=500$, $\delta=30$ days). Full confusion matrices are provided in the supplementary material.}
\begin{tabular}{lcc}
\toprule
Noise level $\sigma$ & Sensitivity (\%) & Specificity (\%) \\
\midrule
$0.25$ & 80.71 $\pm$ 3.13 & 100.0 $\pm$ 0.00 \\
$1.0$  & 82.91 $\pm$ 2.99 & 94.90 $\pm$ 1.28 \\
$2.5$  & 76.45 $\pm$ 3.26 & 82.56 $\pm$ 1.93 \\
$5.0$  &  67.04 $\pm$ 3.31& 86.15 $\pm$ 1.88 \\
\bottomrule
\end{tabular}
\end{table}

\medskip

\textbf{Scenario~\rom{3}} (cohort size)

\begin{figure}[ht]
\centering
    \includegraphics[scale=0.25]{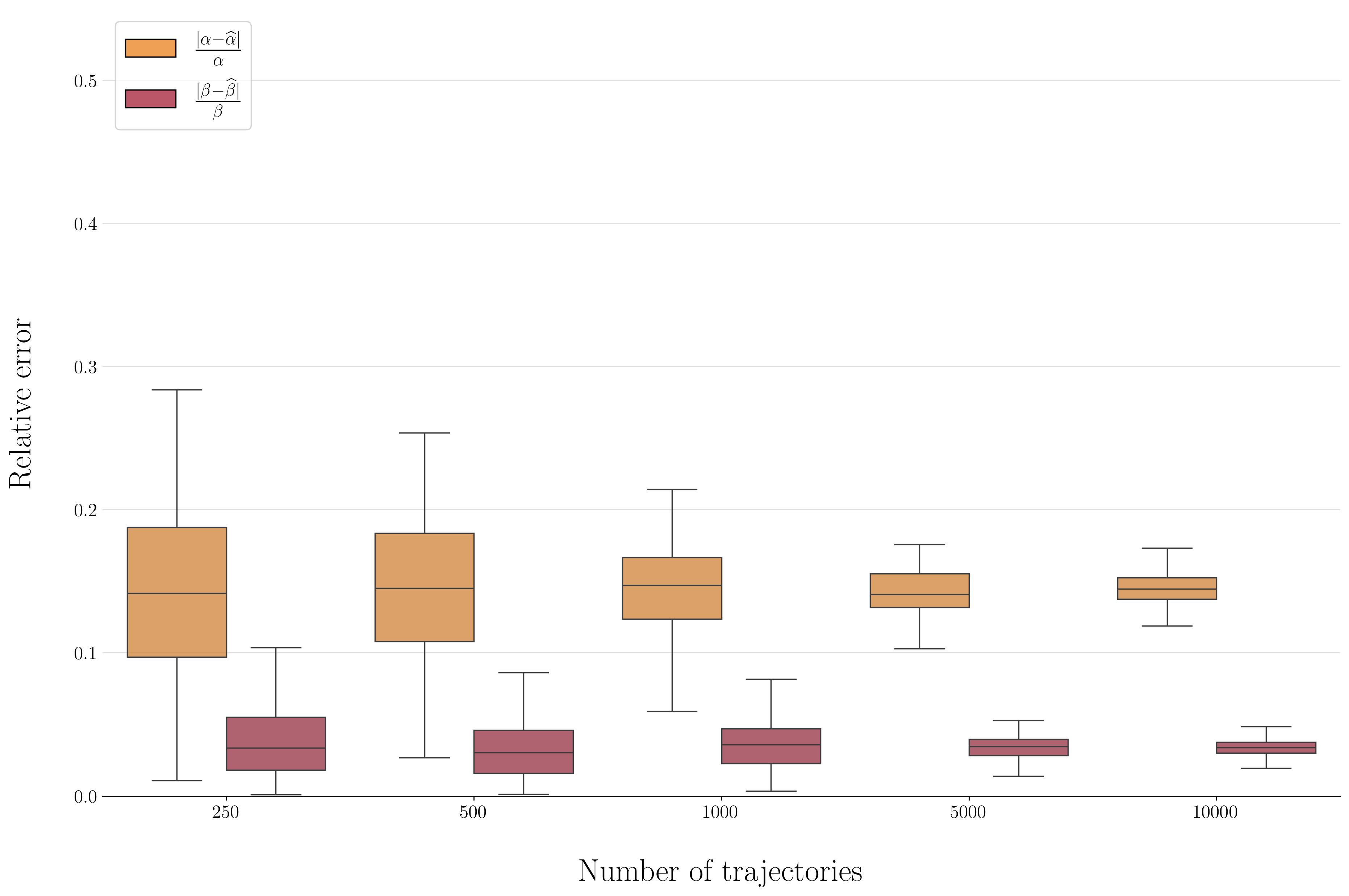}
    \caption{\label{fig:weibullparamscenario3} \textbf{Scenario~III.} Relative errors on Weibull shape ($\alpha$) and scale ($\beta$) parameter estimates across $100$ batches as a function of cohort size $n$ ($\delta=30$ days, $\sigma=1$). Variance decreases with $n$ as expected.}
\end{figure}

Increasing the sample size can be expected to have no impact on the quality of the estimates, and this is borne out in practice (see Supplementary materials, Scenario~\rom{3}). We can nevertheless verify on Figure~\ref{fig:weibullparamscenario3} that the variance of the estimates of the parameters of the survival time distribution decreases with the size of the cohort, as expected.

\medskip

\amelie{\textbf{Scenario~\rom{4}}} (cohort heterogeneity)

\amelie{Using random values for the slopes and visit intervals in each trajectory can be expected to have no impact on the quality of the estimates, and this is borne out in practice (see Supplementary materials, Scenario~\rom{4}). We can nevertheless verify on Figure~\ref{fig:s1s4} that the estimates of the jump parameters are similar between a scenario where $\delta$, $v_{-1}$ and $v_1$ are fixed, and a scenario where they are random. 

\begin{figure}[ht]
\centering
    \includegraphics[scale=0.25]{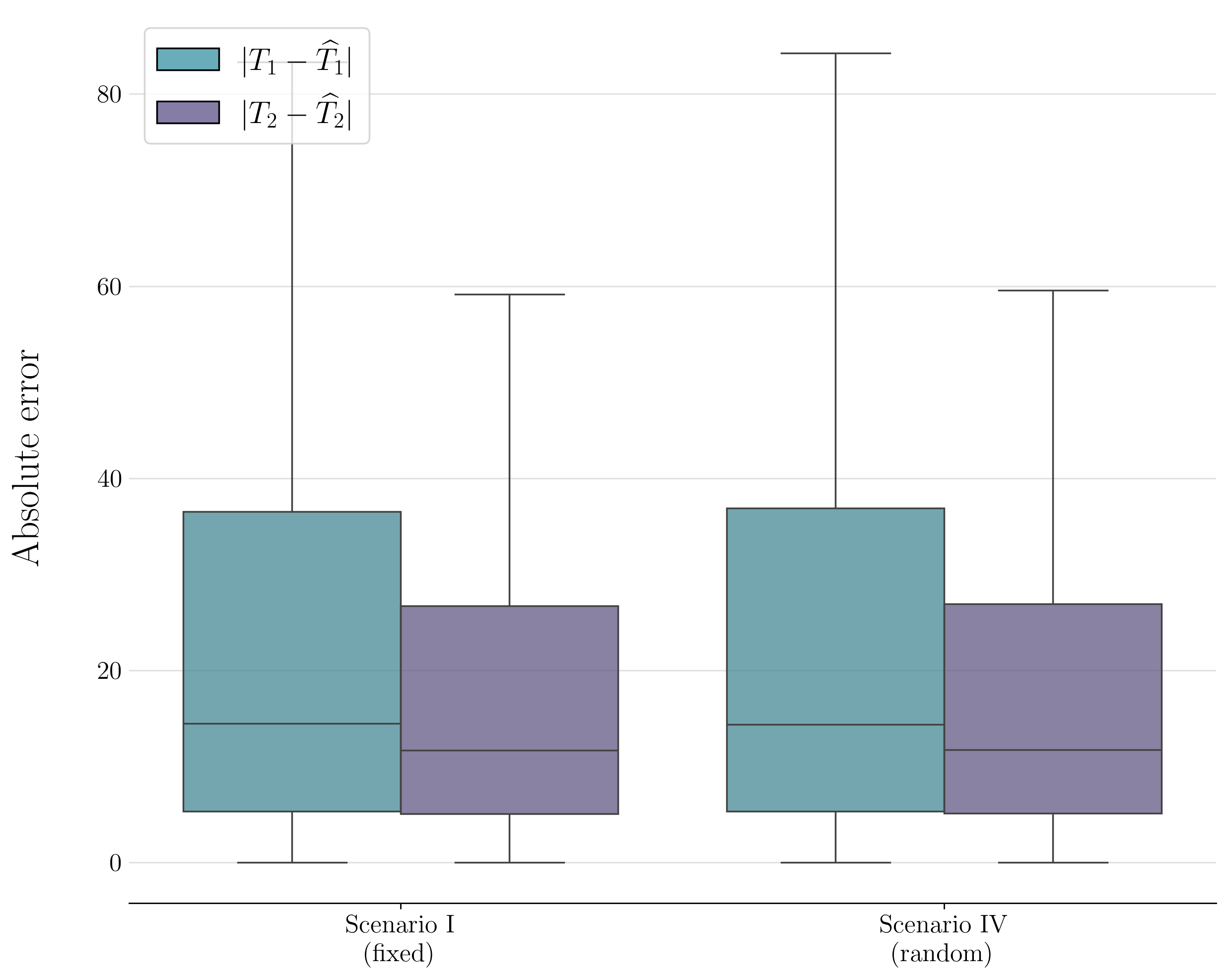}
    \caption{\label{fig:s1s4} \textbf{Scenarios~I and~IV.} Jump time estimation errors (one batch, $n=500$) under fixed parameters (Scenario~I: $\delta=30$\,days, fixed slopes) versus random parameters (Scenario~IV: $\delta\sim\mathcal{U}\{10,\ldots,60\}$, $v_{-1}$ and $v_1$ drawn uniformly). Errors computed only on correctly detected relapses.}
\end{figure}
}

\medskip

\textbf{Discussion on the simulation study}

As already mentioned, our method has an intrinsic bias due to the presence of censoring. We tend to overestimate relapse times, whatever the visit interval or noise level. Further experiments confirm that removing the estimation error on both jump times does alleviate the problem (see Figure~\ref{supp-fig:noesterr} in the Supplementary material). Increasing the follow-up time could improve the estimation of the second jump time, and therefore estimation of the overall the relapse time but we have chosen not to perform such experiment, as this approach would not be realistic for our application case. Reducing the estimation error on relapse times in the presence of censoring remains an open question.

\medskip

\amelie{
\textbf{Extensions}

All the trajectories associated with the results presented here were simulated with the same pair $(\alpha,\beta)$. The same study was carried out with different values of these parameters and an alternative gamma distribution. The results are similar to those of the main scenarios. These are available in Section~\ref{supp-sec:scenariosa} of the Supplementary materials. 
It shows that our approach performs as well in a wide range of settings.

}

\subsection{Comparison with other methods}\label{sec:comparisons}

We now compare our estimation method with \amelie{four alternative methods described below, namely} change point detection, Hidden Markov Models (HMMs) \amelie{threshold and piecewise regression}. \amelie{By default, these} methods exploit little or no information about the model, which is one of their advantages. However in our context,  we choose to give them as much model knowledge as possible to improve their performance. We develop the estimation procedure with change point detection, HMMs, \amelie{threshold and piecewise regression}, and explain how we compare them with our method.

\subsubsection{Change point detection}

Given a non-stationary signal $y$, (offline) change point detection aims to find the best segmentation of $y$ into a piecewise stationary signal by detecting when the signal changes dynamics. This is done by minimizing some predefined criterion.

The signal considered is the logarithm of the consecutive differences of the simulated trajectories. That is, if $(y_i)_{i=1,\ldots,N}$ is an $N$-length trajectory, the signal is $(\mathrm{log}(y_i) - \mathrm{log}(y_{i+1}))_{i=1,\ldots,N-1}$. We focus on trajectories where a relapse occurs and therefore set the number $K$ of breakpoints to be detected at  $K=2$. This allows us to recover the $3$ modes of the process. We assume that our signal is Gaussian, with piecewise constant mean and fixed variance. We thus use a quadratic error loss as criterion function, a common practice in such cases.

Change point detection is implemented using the \texttt{ruptures} python library \citep{truong2020review}.

\subsubsection{HMMs}

Here our trajectories are seen as a sequence of observations linked to hidden states of an underlying Markov process. We use an HMM to recover these hidden states, which correspond to the modes of the trajectories. An HMM is determined by the transition probabilities between states, the parameters of the emission probability distributions and the initial state distribution. Given the observations, we seek to estimate the model parameters and infer the hidden states.

Again, the sequence of observations is the logarithm of consecutive differences from the trajectories. We use model information to initialize and constrain some of the model parameters. First, as with change point detection we only consider trajectories where a relapse truly occurs and fix the number of hidden states to $3$. Then, we initiate the initial state distribution so that the probability of starting in mode $m=-1$ is $1$, according to our simulation procedure. The initial transition probability matrix $\mathbf{A}$ is given in Equation (\ref{eq:transmat}). For example, given that the process is currently in mode $m=-1$, the probability of switching to mode $m=0$ is initiated to $0.7$, hence higher than staying in mode $m=0$. This broadly reflects the behaviour of the trajectories. Finally, we constrain the transition matrix to prevent impossible transitions and fix the absorbing state. Denoting $m_t$ the mode at time $t$, the transition matrix is such that $\mathbb{P}(m_t=-1 \mid m_{t-1}=0) = \mathbb{P}(m_t=-1 \mid m_{t-1}=1) = 0$, $\mathbb{P}(m_t=0 \mid m_{t-1}=1) = 0$, $\mathbb{P}(m_t=1 \mid m_{t-1}=-1) = 0$ and $\mathbb{P}(m_t=1 \mid m_{t-1}=1) = 1$:

\begin{equation}\label{eq:transmat}
    \mathbf{A} = \begin{blockarray}{cccr}
        {\color{Maroon}-1} & {\color{Maroon}0} & {\color{Maroon}1} \\
        \begin{block}{(ccc)r}
            0.3 & 0.7 & 0.0 & {\color{Maroon}-1} \\
            0.0 & 0.4 & 0.6 & {\color{Maroon}0} \\
            0.0 & 0.0 & 1.0 & {\color{Maroon}1} \\
        \end{block}
    \end{blockarray}
\end{equation}

The model parameters are then estimated using an adapted version of the \texttt{hmmlearn} python package \citep{hmmlearn}.

\amelie{
\subsubsection{Threshold}

This method estimates remission and relapse times as the midpoints of the intervals defined by the visits immediately before and after crossing a fixed threshold $\zeta_{\text{th}}$. For the relapse time $T_2$, the trajectory is treated in reverse. If the last point on a trajectory is below the threshold, it is considered that there has been no relapse and the trajectory is censored. We consider three different thresholds, namely $2$, $5$ and $10$.
}

\amelie{
\subsubsection{Piecewise regression}

Piecewise regression aims to partition a signal by identifying abrupt changes, known as breakpoints, and fitting a regression model to data given those breakpoints.
The signal considered is the logarithm of the simulated trajectories and the breakpoints correspond to the jump times $T_1$ and $T_2$. We perform piecewise regression with $1$ and $2$ breakpoints on each trajectory and keep the best model in terms of BIC. If there is only $1$ breakpoint, it is considered that there has been no relapse and the trajectory is censored.

Piecewise regression is implemented using the \texttt{piecewise-regression} python library \citep{pilgrim2021piecewise}.

}

\subsubsection{Performance comparison}

All five methods are evaluated on a common set of simulated trajectories. The five methods differ in how they determine whether a relapse has occurred (model selection / censoring criterion):
\begin{itemize}
\item \textbf{PDMP (ours)}: a subject is declared censored if the estimated relapse slope satisfies $\widehat{v_1} < v_{\min} = 0.005$, or if the 5 most recent post-remission observations are all below $3\,\si{\micro\gram\per\litre}$ (see Section~\ref{subsubsec:jump time estimation}).
\item \textbf{Threshold}: a subject is declared censored if the last observation of the trajectory is below the threshold $\zeta_{\text{th}}$.
\item \textbf{Piecewise regression}: BIC is used to select between a model with one breakpoint (censored) and a model with two breakpoints (relapse detected).
\item \textbf{HMM and change point}: We tried to let both methods deduce the number of mode changes using common model selection heuristics (including BIC, ICL and other in-house criteria), but the results were very poor. We therefore parametrized these methods to  always predict two mode transitions and produce no censoring indicator.
\end{itemize}

\medskip

\begin{figure}[htbp]
\centering
    \includegraphics[width=\textwidth]{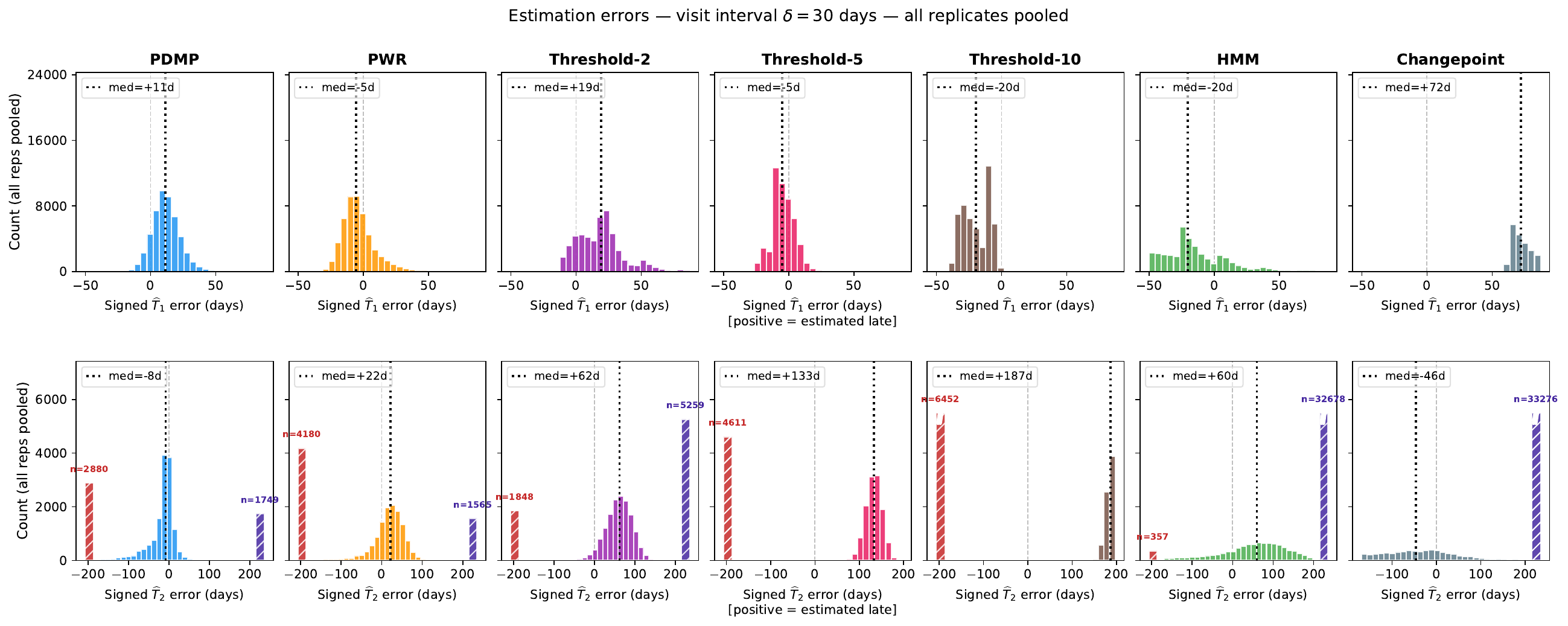}
    \caption{\label{fig:errorhist}
    Signed estimation errors for $\widehat{T_1}$ (top row) and $\widehat{T_2}$ (bottom row)
    for all methods, at visit interval $\delta=30$ days, pooled across all simulation
    replicates. Positive values indicate the event was estimated later than its true time;
    the dotted vertical line marks the median error over correctly identified events.
    For $\widehat{T_2}$, hatched bars at the edges of the x-axis indicate
    misclassifications: crimson (left) for missed relapses (false censoring) and
    indigo (right) for spurious relapses (false relapse); bars are capped at the
    y-axis scale and annotated with the actual count.
    }
\end{figure}

\begin{figure}[htbp]
\centering
    \includegraphics[width=0.75\textwidth]{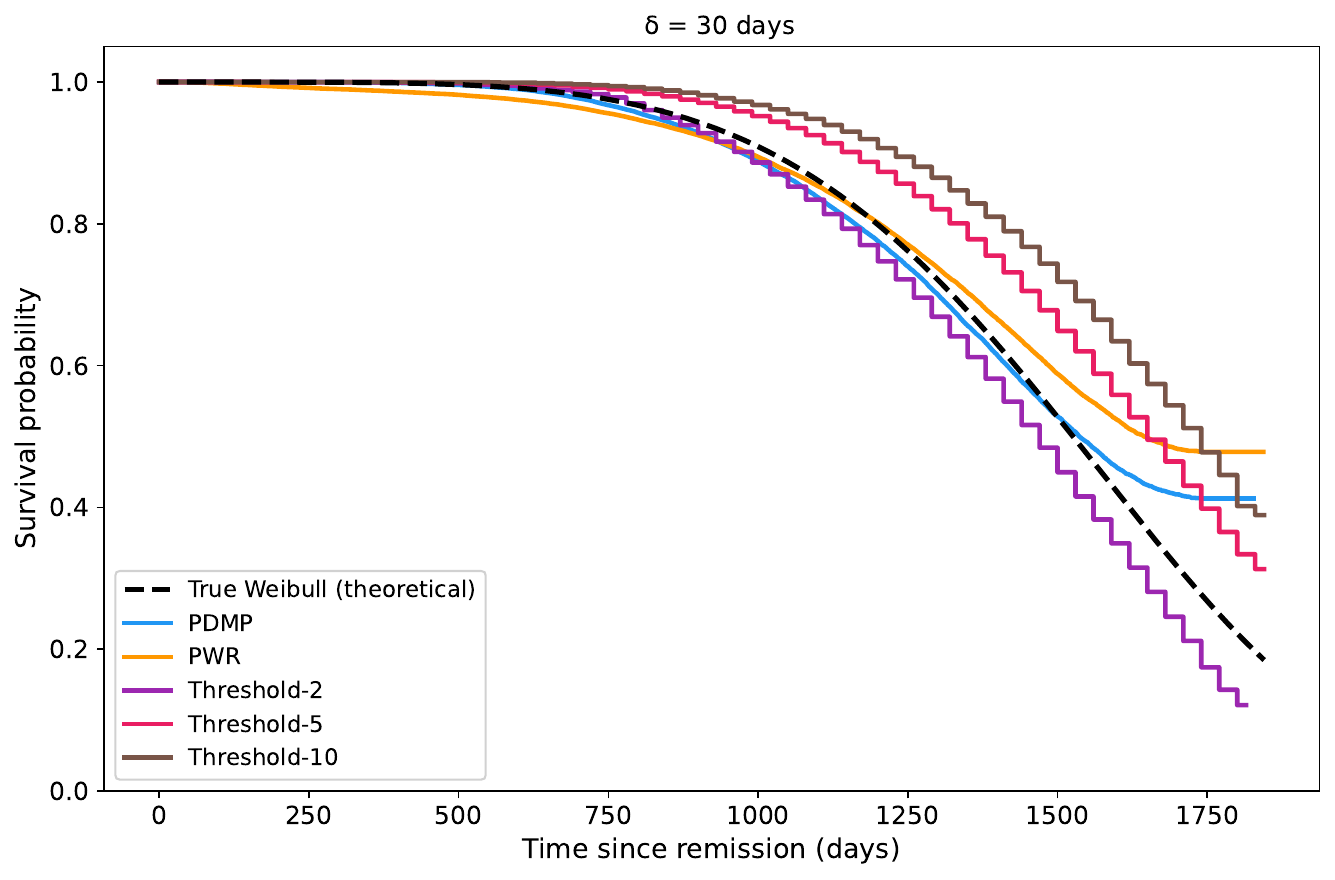}
    \caption{\label{fig:kmcomparison}
    Kaplan-Meier survival curves estimated by each method (excluding HMM and
    change point, which do not produce a valid censoring indicator), alongside
    the true Weibull survival function (dashed). 
    The integrated $L_1$ distance between each method's KM curve and the true
    Weibull function is reported in Table~\ref{tab:comparison_delta30}.
    }
\end{figure}

\amelie{
\textbf{Unified comparison framework}\quad
We compare all methods using the following metrics: (i)~\textit{sensitivity} (proportion of true relapses correctly detected), (ii)~\textit{specificity} (proportion of truly censored subjects correctly identified), (iii)~\textit{Adjusted Rand Index} (ARI) measuring partition accuracy on all trajectories, (iv)~\textit{integrated $L_1$ distance} between the estimated Kaplan--Meier survival curve and the true Weibull survival function (a global goodness-of-fit measure; lower is better), and (v)~\textit{compute time}. Results are summarised in Table~\ref{tab:comparison_delta30} for the primary visit interval $\delta=30$ days. Figure~\ref{fig:errorhist} shows the signed estimation errors for $\widehat{T_1}$ and $\widehat{T_2}$ for all methods at $\delta=30$ (hatched bars at the histogram extremes indicate misclassifications: crimson for missed relapses, indigo for spurious detections). Figure~\ref{fig:kmcomparison} shows the Kaplan--Meier survival curves estimated by each method alongside the true Weibull. Additional results for $\delta \in \{10, 60\}$ days are provided in Tables~\ref{tab:comparison_delta10} and~\ref{tab:comparison_delta60}. For each value of $\delta$, 100 replicates of $n=500$ patient trajectories were simulated. 
}

\smallskip

\begin{table}[htbp]
\centering
\caption{\label{tab:comparison_delta30}%
Performance comparison of all methods — visit interval $\delta=30$ days
($n=500$ trajectories per replicate).
Sensitivity and specificity (\%) measure relapse detection accuracy.
ARI (Adjusted Rand Index) measures mode-partition accuracy on all trajectories.
$L_1$ distance (days) measures the integrated absolute deviation between
the estimated and true Weibull survival curves (lower is better).
$^\dagger$ HMM and change point do not produce a censoring indicator;
their $L_1$ values are computed under the assumption that all subjects
relapsed and should be interpreted with caution.
}
\resizebox{\textwidth}{!}{%
\begin{tabular}{lccccr}
\toprule
Method & Sensitivity (\%) & Specificity (\%) & ARI & $L_1$ distance (days) & Compute time \\
\midrule
\textbf{PDMP (ours)} & 82.8 $\pm$ 3.0 & 94.7 $\pm$ 1.2 & \textbf{0.881 $\pm$ 0.006} & \textbf{42.731 $\pm$ 10.481} & 2.2\,s \\
Piecewise regression & 75.0 $\pm$ 3.2 & 95.3 $\pm$ 1.3 & 0.881 $\pm$ 0.008 & 73.653 $\pm$ 13.086 & 1.7\,min \\
Threshold ($\zeta_{\mathrm{th}}=2$) & 89.0 $\pm$ 2.5 & 84.2 $\pm$ 2.1 & 0.826 $\pm$ 0.006 & 47.244 $\pm$ 13.435 & 3.2\,ms \\
Threshold ($\zeta_{\mathrm{th}}=5$) & 72.4 $\pm$ 3.6 & \textbf{100.0 $\pm$ 0.0} & 0.863 $\pm$ 0.007 & 98.935 $\pm$ 16.046 & 3.0\,ms \\
Threshold ($\zeta_{\mathrm{th}}=10$) & 61.4 $\pm$ 4.0 & \textbf{100.0 $\pm$ 0.0} & 0.738 $\pm$ 0.008 & 149.309 $\pm$ 17.857 & 2.8\,ms \\
HMM$^\dagger$ & 97.9 $\pm$ 1.1 & 1.8 $\pm$ 0.8 & 0.648 $\pm$ 0.015 & 1189.129 $\pm$ 33.423 & 4.1\,s \\
Change point$^\dagger$ & \textbf{100.0 $\pm$ 0.0} & 0.0 $\pm$ 0.0 & 0.179 $\pm$ 0.010 & 1011.463 $\pm$ 36.183 & 181.4\,ms \\
\bottomrule
\end{tabular}%
}
\end{table}

\begin{table}[htbp]
\centering
\caption{\label{tab:comparison_delta10}%
Performance comparison of all methods — visit interval $\delta=10$ days
($n=500$ trajectories per replicate).
Sensitivity and specificity (\%) measure relapse detection accuracy.
ARI (Adjusted Rand Index) measures mode-partition accuracy on all trajectories.
$L_1$ distance (days) measures the integrated absolute deviation between
the estimated and true Weibull survival curves (lower is better).
$^\dagger$ HMM and change point do not produce a censoring indicator;
their $L_1$ values are computed under the assumption that all subjects
relapsed and should be interpreted with caution.
}
\resizebox{\textwidth}{!}{%
\begin{tabular}{lccccr}
\toprule
Method & Sensitivity (\%) & Specificity (\%) & ARI & $L_1$ distance (days) & Compute time \\
\midrule
\textbf{PDMP (ours)} & 86.6 $\pm$ 2.7 & 91.8 $\pm$ 1.7 & \textbf{0.934 $\pm$ 0.004} & 31.019 $\pm$ 9.757 & 2.2\,s \\
Piecewise regression & 77.9 $\pm$ 3.1 & 97.7 $\pm$ 0.8 & 0.901 $\pm$ 0.005 & 65.777 $\pm$ 15.308 & 1.7\,min \\
Threshold ($\zeta_{\mathrm{th}}=2$) & 88.7 $\pm$ 2.4 & 83.9 $\pm$ 2.1 & 0.863 $\pm$ 0.004 & \textbf{29.282 $\pm$ 10.318} & 3.2\,ms \\
Threshold ($\zeta_{\mathrm{th}}=5$) & 73.5 $\pm$ 3.3 & \textbf{100.0 $\pm$ 0.0} & 0.705 $\pm$ 0.005 & 116.180 $\pm$ 15.767 & 3.0\,ms \\
Threshold ($\zeta_{\mathrm{th}}=10$) & 61.9 $\pm$ 4.0 & \textbf{100.0 $\pm$ 0.0} & 0.545 $\pm$ 0.005 & 167.414 $\pm$ 15.816 & 2.8\,ms \\
HMM$^\dagger$ & 98.8 $\pm$ 0.8 & 4.0 $\pm$ 1.1 & 0.682 $\pm$ 0.010 & 1118.986 $\pm$ 23.387 & 4.1\,s \\
Change point$^\dagger$ & \textbf{100.0 $\pm$ 0.0} & 0.0 $\pm$ 0.0 & 0.225 $\pm$ 0.009 & 1292.013 $\pm$ 27.559 & 181.4\,ms \\
\bottomrule
\end{tabular}%
}
\end{table}

\begin{table}[htbp]
\centering
\caption{\label{tab:comparison_delta60}%
Performance comparison of all methods — visit interval $\delta=60$ days
($n=500$ trajectories per replicate).
Sensitivity and specificity (\%) measure relapse detection accuracy.
ARI (Adjusted Rand Index) measures mode-partition accuracy on all trajectories.
$L_1$ distance (days) measures the integrated absolute deviation between
the estimated and true Weibull survival curves (lower is better).
$^\dagger$ HMM and change point do not produce a censoring indicator;
their $L_1$ values are computed under the assumption that all subjects
relapsed and should be interpreted with caution.
}
\resizebox{\textwidth}{!}{%
\begin{tabular}{lccccr}
\toprule
Method & Sensitivity (\%) & Specificity (\%) & ARI & $L_1$ distance (days) & Compute time \\
\midrule
\textbf{PDMP (ours)} & 79.2 $\pm$ 3.3 & 96.2 $\pm$ 1.1 & 0.770 $\pm$ 0.006 & \textbf{60.502 $\pm$ 9.597} & 2.2\,s \\
Piecewise regression & 65.3 $\pm$ 3.4 & 85.9 $\pm$ 2.0 & 0.811 $\pm$ 0.012 & 112.325 $\pm$ 15.472 & 1.7\,min \\
Threshold ($\zeta_{\mathrm{th}}=2$) & 88.2 $\pm$ 2.5 & 84.2 $\pm$ 1.9 & 0.773 $\pm$ 0.008 & 76.654 $\pm$ 16.899 & 3.2\,ms \\
Threshold ($\zeta_{\mathrm{th}}=5$) & 72.2 $\pm$ 3.1 & \textbf{100.0 $\pm$ 0.0} & \textbf{0.888 $\pm$ 0.007} & 80.238 $\pm$ 15.161 & 3.0\,ms \\
Threshold ($\zeta_{\mathrm{th}}=10$) & 61.5 $\pm$ 3.4 & \textbf{100.0 $\pm$ 0.0} & 0.861 $\pm$ 0.008 & 115.311 $\pm$ 13.589 & 2.8\,ms \\
HMM$^\dagger$ & 99.0 $\pm$ 0.8 & 0.9 $\pm$ 0.5 & 0.599 $\pm$ 0.016 & 1157.436 $\pm$ 42.519 & 4.1\,s \\
Change point$^\dagger$ & \textbf{100.0 $\pm$ 0.0} & 0.0 $\pm$ 0.0 & 0.129 $\pm$ 0.007 & 663.484 $\pm$ 11.003 & 181.4\,ms \\
\bottomrule
\end{tabular}%
}
\end{table}


  \textbf{Adjusted Rand Index, L1 metric (Tables~\ref{tab:comparison_delta30} to~\ref{tab:comparison_delta60})}

  Our method achieves the highest ARI and lowest variability across all trajectories (Table~\ref{tab:comparison_delta30}) as well as the best trade-off between specificity and sensitivity, with almost equivalent performances of piecewise regression and the threshold methods with low threshold. The gap between the methods widens as the visits are closer (Table~\ref{tab:comparison_delta10}). Importantly, the PDMP approach provides the best results in terms of reconstructing the true relapse time distribution, as measured by the L1 distance criteria. The gap between HMM and change point is also noteworthy: the probabilistic emission model of the HMM provides better partition accuracy than the purely signal-based change point approach, even though neither method can identify censored subjects. This confirms that exploiting model knowledge, even partially, yields substantially better trajectory partitioning.

\medskip

\textbf{Signed estimation errors (Figure~\ref{fig:errorhist})}

The error histograms reveal the characteristic biases of each method.
For~$\widehat{T_1}$, PDMP has a slight positive median bias ($+11$ days),
reflecting that the algorithm requires a few post-threshold observations to
confirm remission. Threshold methods behave predictably: a higher threshold
produces earlier $\widehat{T_1}$ estimates (increasingly negative bias) since
the descending biomarker crosses a high threshold before reaching the lower
remission level $\zeta_r$. Change point produces highly inaccurate $T_1$
estimates (median $+72$ days), illustrating the difficulty of purely
signal-based methods with smooth exponential trajectories.

For~$\widehat{T_2}$, PDMP achieves near-zero median error ($-8$ days) with a compact distribution, consistent with its performance in $L_1$ distance.
The sentinel bars show $2{,}880$ missed relapses and $1{,}749$ false relapses (out of $50000$ trajectories),
reflecting the $83\%/95\%$ sensitivity/specificity trade-off. 
PWR exhibits systematic late bias ($+22$ days), which inflates its
estimated survival times. The threshold methods display a monotone pattern:
higher thresholds produce far more missed relapses (left sentinel, crimson) and
fewer false relapses (right sentinel, indigo), together with increasingly large
positive $T_2$ bias. HMM and change point, which never declare censoring, are
dominated by false relapses ($n>32{,}000$ each), confirming they are
fundamentally unsuitable for censoring classification despite competitive $T_1$
accuracy (HMM, median $-20$ days).

\smallskip
\textbf{Kaplan--Meier survival curves (Figure~\ref{fig:kmcomparison})}

The KM curves translate per-subject errors into a global view of survival
distribution recovery. PDMP tracks the true Weibull curve most closely
throughout, with only a slight upward deviation at long follow-up times due
to residual over-censoring. PWR lies consistently above the true curve,
a direct consequence of its positive $T_2$ bias.
Among the threshold methods, Threshold-2 lies closest to the true curve
despite its high false positive rate: its two biases
(false positives pulling the curve down, missed relapses pulling it up)
partially cancel, yielding a competitive $L_1$ of $47$\,days compared to
PDMP's $43$\,days. Threshold-5 and Threshold-10 lie progressively further
above the true curve, reflecting their increasing missed-relapse rates
($27\%$ and $38\%$ respectively).
The $L_1$ ranking ($\text{PDMP} < \text{Threshold-2} < \text{PWR} <
\text{Threshold-5} < \text{Threshold-10}$) is directly visible in the figure:
the area between each method's curve and the dashed true curve corresponds
exactly to its $L_1$ value.

\smallskip
\textbf{Conclusion of the comparison}

For all alternative methods, the presence of censoring requires a model selection procedure of some sort prior to recovering the relapse time. The comparison conducted in this section has shown that although our method is in essence similar to piecewise regression, separating the estimation of $T_1$ from that of $T_2$ and the censoring indicator produces better results than a joint estimation of these parameters in terms of estimation error. Our method is also considerably faster in terms of computing time. We note that piecewise regression could in principle adopt our censoring criterion (slope $\widehat{v_1} < v_{\min}$ and end-of-follow-up check) in place of BIC model selection, which might improve its sensitivity without degrading specificity; we leave this variant for future work. The key advantage of our sequential approach over piecewise regression is that our censoring criterion ($\widehat{v_1} < 0.005$) is applied to a residual trajectory already conditioned on $\widehat{T_1}$, which removes the influence of the first phase and reduces the variability of the relapse slope estimate. By contrast, piecewise regression performs a global joint fit of both breakpoints simultaneously: when the position of $T_1$ is imprecise, the global optimisation displaces $T_2$ to compensate, inflating its estimation error even on trajectories where relapse is ultimately detected. Our sequential decomposition breaks this dependency, isolating the $T_2$ estimation problem from errors in $T_1$.

\section{Application to myeloma}\label{sec:MM}

\begin{figure}[ht]
  \captionsetup[sub]{labelformat=empty, position=top}
  \centering
  \subcaptionbox[b]{\label{fig:vremission}}{\includegraphics[width=0.45\textwidth]{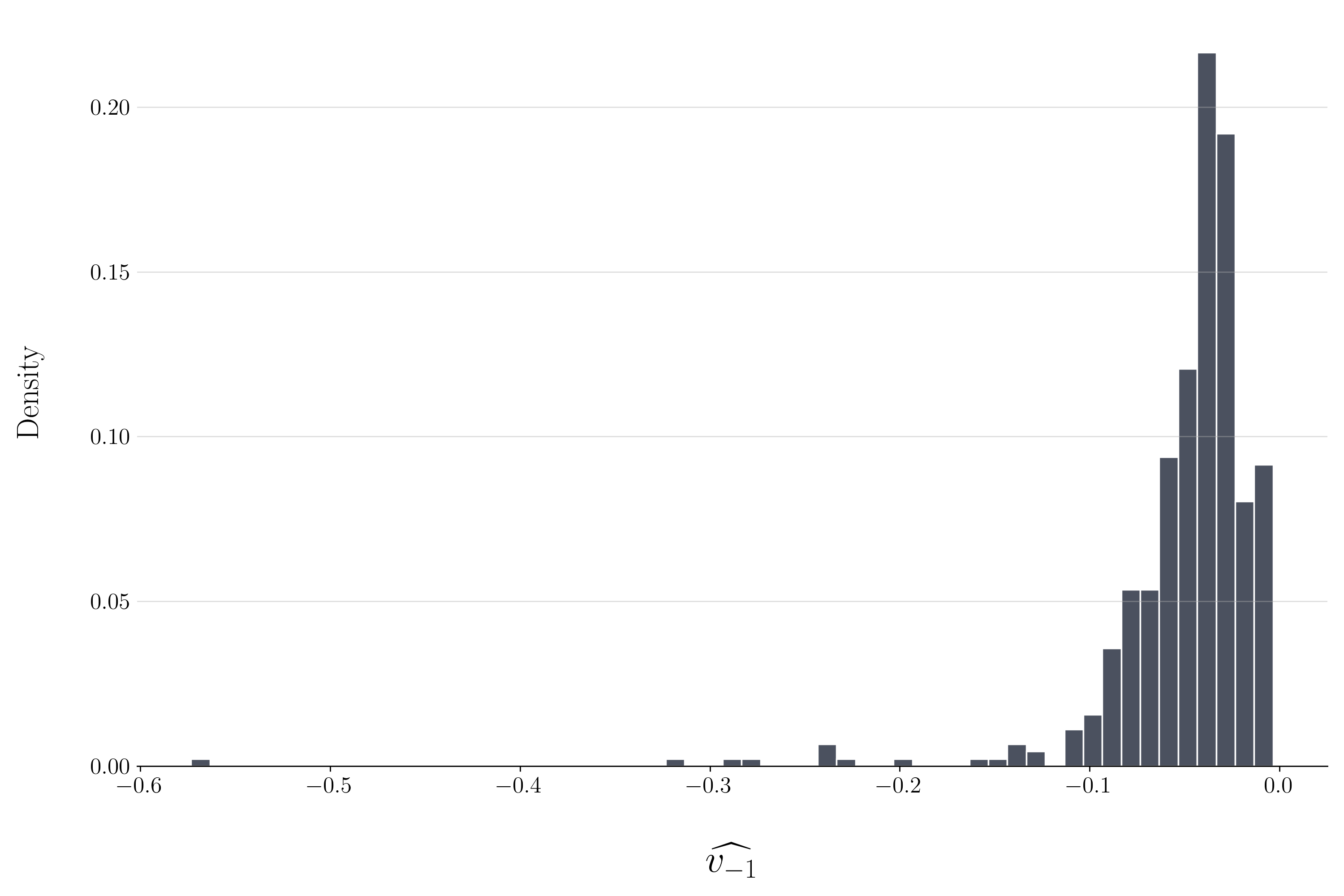}}\hspace{1ex}%
  \subcaptionbox[b]{\label{fig:vrelapse}}{\includegraphics[width=0.45\textwidth]{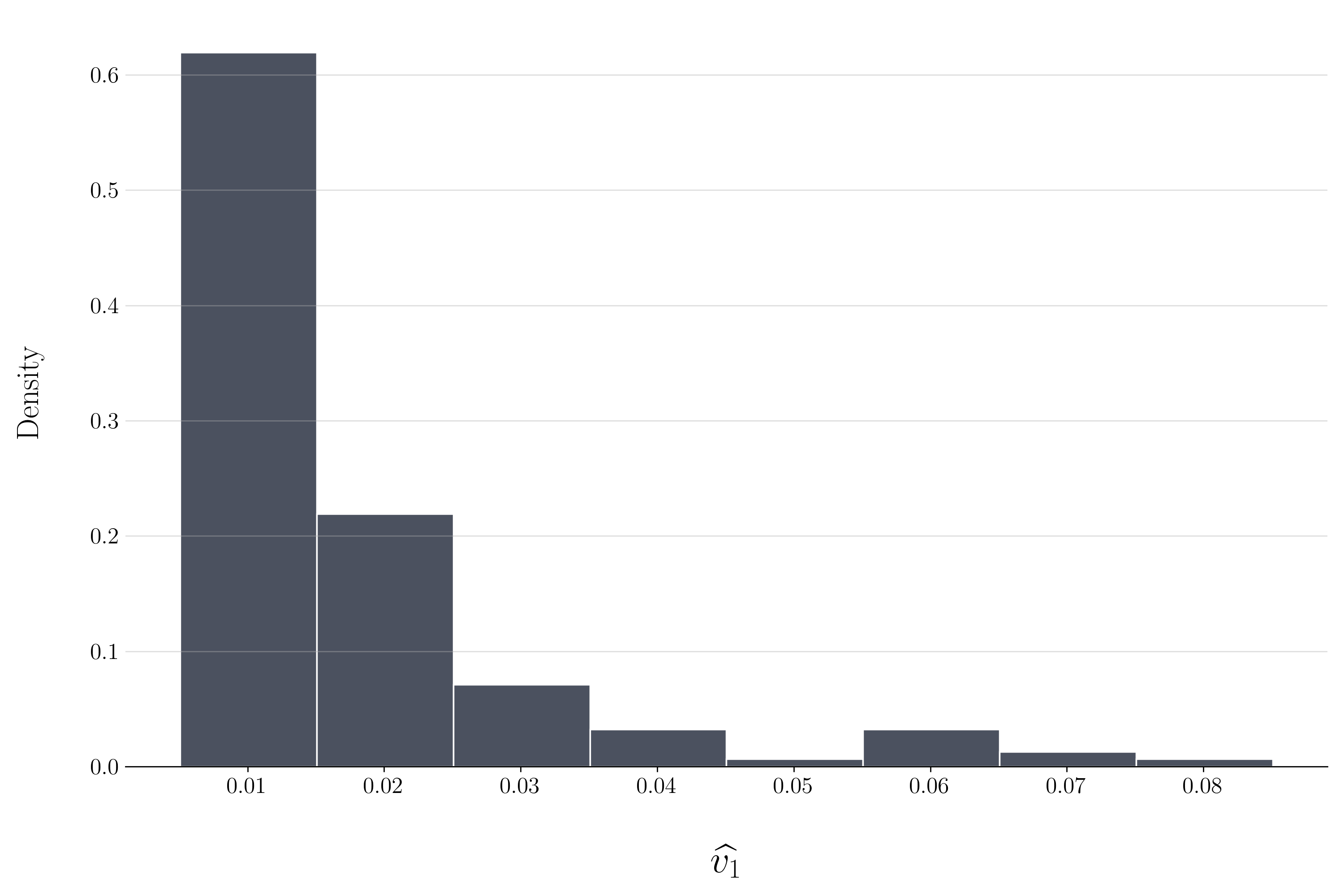}}
  \caption{\label{fig:slopeshist} Distribution of the estimated remission slopes $v_{-1}$ and relapse slopes $v_{1}$ in the application dataset.}
\end{figure}

\begin{figure}[htbp]
\centering
    \includegraphics[width=0.48\textwidth]{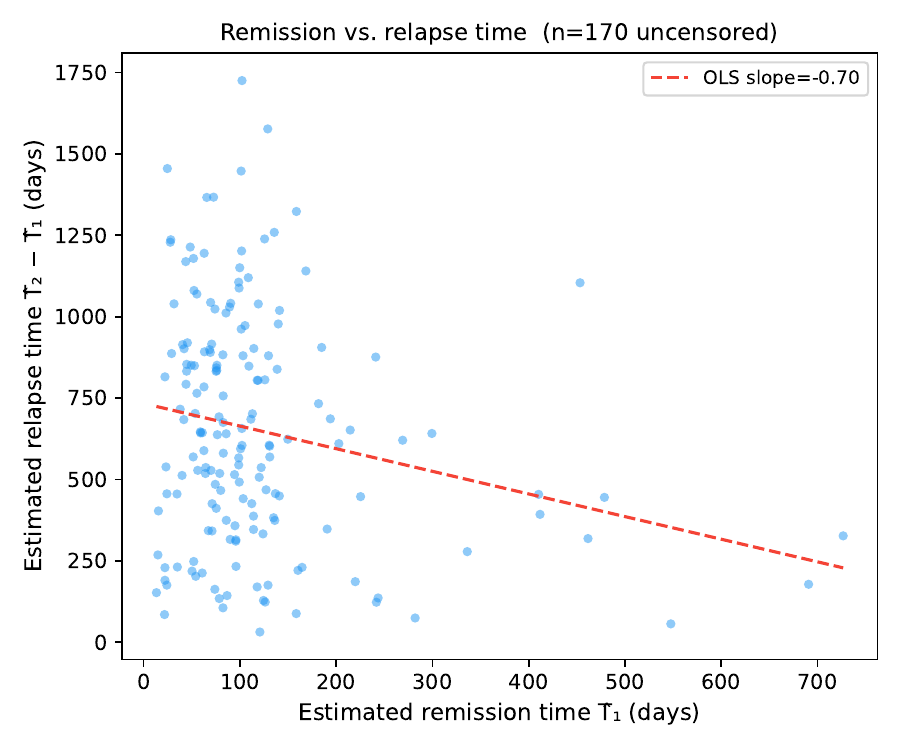}
    \hfill
    \includegraphics[width=0.48\textwidth]{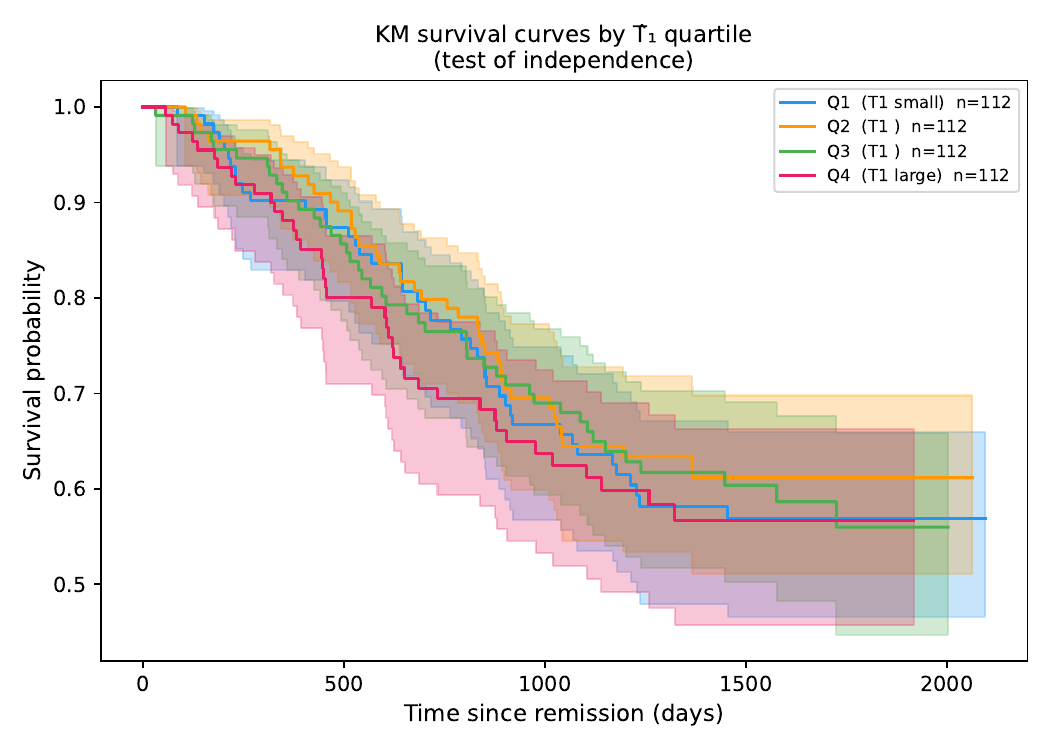}
    \caption{\label{fig:independence}
    Assessment of the independence between $\widehat{T_1}$ and $\widehat{T_2}-\widehat{T_1}$
    on the real myeloma data ($n=170$ subjects with a detected relapse out of $448$).
    \textit{Left}: scatter plot of estimated remission time $\widehat{T_1}$ versus
    estimated relapse duration $\widehat{T_2}-\widehat{T_1}$, with a linear regression
    line (Spearman $\rho=-0.15$, $p=0.053$; Kendall $\tau=-0.10$, $p=0.044$).
    \textit{Right}: Kaplan-Meier survival curves of $\widehat{T_2}-\widehat{T_1}$
    stratified by quartile of $\widehat{T_1}$; a multivariate log-rank test across
    the four strata yields $p<0.001$, suggesting that independence may only hold
    approximately in the real data.
    }
\end{figure}

\begin{figure}
    \centering
    \includegraphics[width=0.5\linewidth]{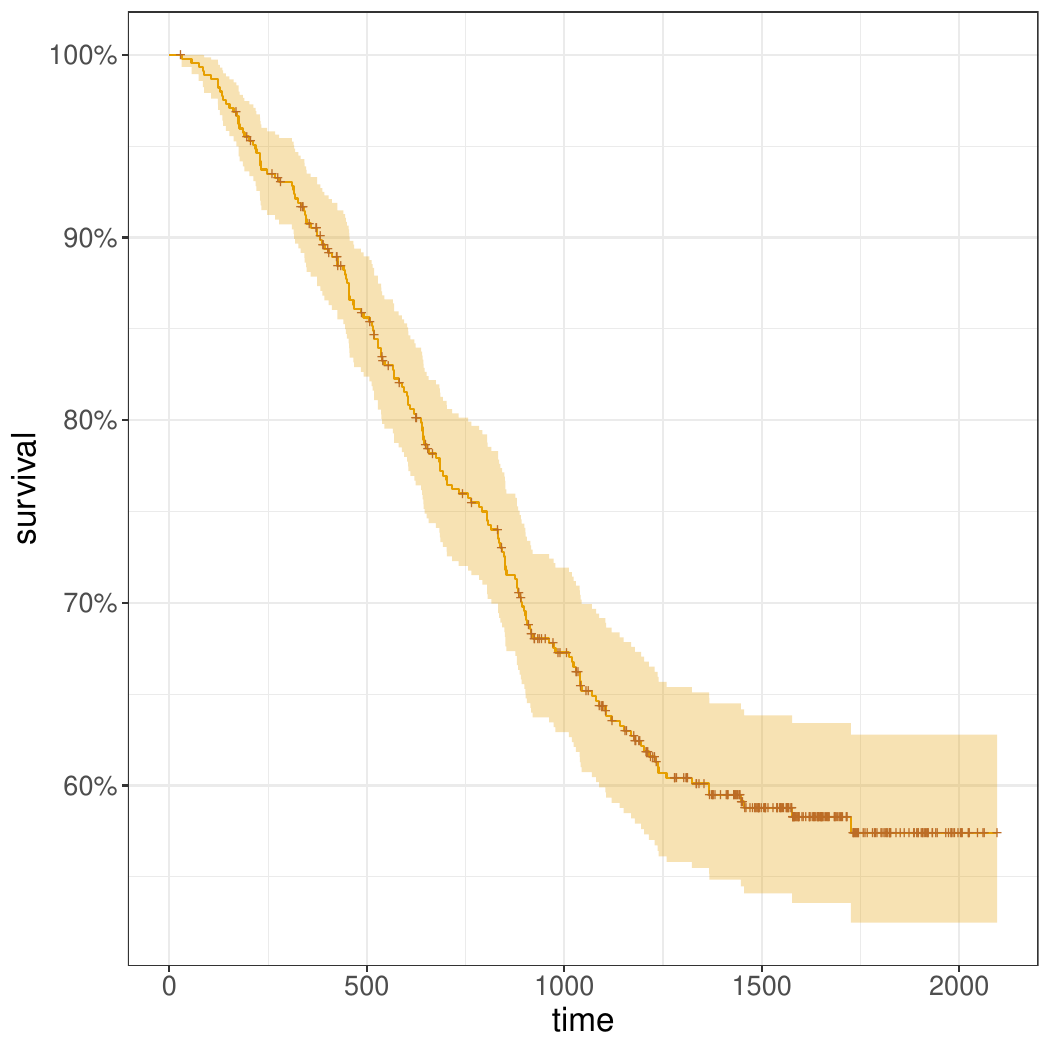}
    \caption{Kaplan-Meier estimates of the probability of survival over time, together with its pointwise $95\%$ confidence intervals. Crosses represent censored events.}
    \label{fig:kmappli}
\end{figure}

\begin{figure}[ht]
\centering
\includegraphics[width=.80\textwidth]{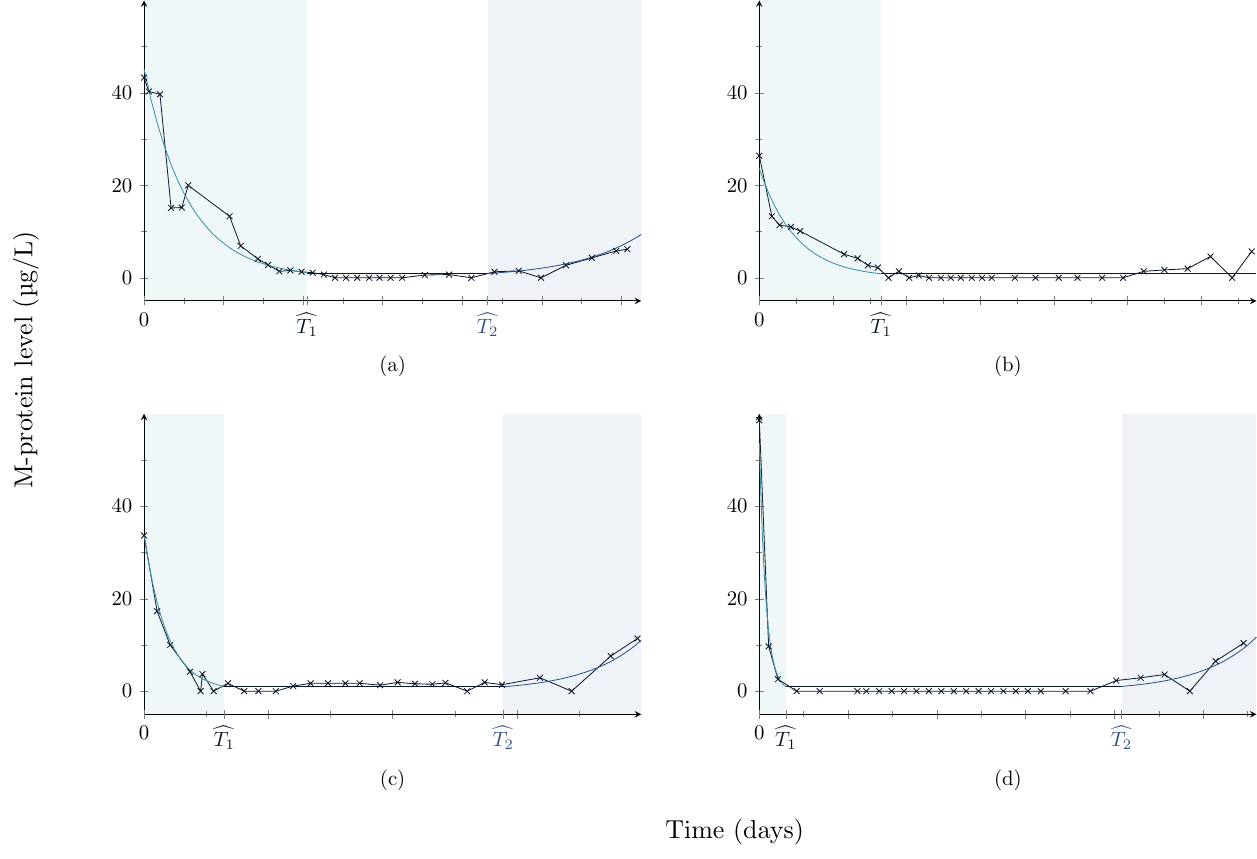}
\caption{\label{fig:trajectories_fitted} Some examples of trajectories from the dataset. The crosses correspond to observations and the black lines in between are interpolations. The estimated states of the patient are represented by the highlighted areas. The patient is considered in mode $m=-1$ between time $t=0$ and $t=\widehat{T_1}$, in mode $m=0$ between $t=\widehat{T_1}$ and $t=\widehat{T_2}$, and in mode $m=1$ after $t=\widehat{T_2}$. The estimated flow is represented by the curve with the corresponding colours.}
\end{figure}

The proposed estimation method is applied to the data of interest. The number of trajectories is $479$, as mentioned in Section~\ref{sec:data}. \amelie{The estimated censoring rate is $62\%$. The Kaplan Meier estimator for the survival time distribution before relapse is represented in Figure~\ref{fig:kmappli}.} The trajectories from the dataset presented in Figure~\ref{fig:trajectories} are shown in Figure~\ref{fig:trajectories_fitted} with the parameters obtained by the estimation procedure. Note that in our simulation study, we used regular time intervals and similar $\widehat{v_{-1}}$ and $\widehat{v_{1}}$ slopes between patients for simplicity, but our method does not rely on these assumptions. In real data, visits are not equally spaced and the method still works. \amelie{The distribution of $\widehat{v_{-1}}$ and $\widehat{v_{1}}$ is given if Figure~\ref{fig:slopeshist}. The estimated remission time $\widehat{T_1}$ ranges from $13$ to $727$, approximately, with a mean value of about $116$. The estimated relapse time $\widehat{T_2}$ ranges from $106$ to $1827$, approximately, with a mean value of about $769$.} Figure~\ref{fig:independence} assesses the independence of $\widehat{T_1}$ and $\widehat{T_2} - \widehat{T_1}$ through a scatter plot and stratified Kaplan-Meier analysis.
The stratified log-rank test and Spearman correlation suggest a violation of the Markov assumption: patients with longer estimated remission times $\widehat{T_1}$ tend to have shorter relapse-free intervals $\widehat{T_2}-\widehat{T_1}$. \\
We believe that this reflects latent patient heterogeneity rather than a structural deficiency of the model. Using the clinical response data available for all $170$ patients with a detected relapse, we partitioned them into two groups: those who achieved Stringent Complete Response at some point during follow-up ($n=61$, 36\%) and those who did not ($n=109$, 64\%). These two groups have markedly different relapse dynamics: SCR patients have a median relapse-free survival of $851$ days, compared to $492$ days for non-SCR patients (Mann-Whitney $p<0.001$). Crucially, within each group the dependence between $\widehat{T_1}$ and $\widehat{T_2}-\widehat{T_1}$ vanishes: (Spearman $\rho=-0.053$ $p=0.69$ for SCR patients and $\rho=-0.115$, $p=0.23$ for non-SCR patients). The marginal correlation observed in the pooled analysis is therefore entirely explained by the confounding effect of response depth: SCR patients achieve remission more rapidly (shorter $\widehat{T_1}$) and remain in remission longer (larger $\widehat{T_2}-\widehat{T_1}$), while non-SCR patients follow the opposite pattern. Within each homogeneous population, the Markov assumption holds. This suggests a natural extension of the model: conditioning the relapse time distribution on depth of response, or fitting a mixture PDMP in which SCR and non-SCR patients follow distinct Weibull distributions.

\medskip

\medskip

\medskip

\section{Discussion and conclusion}

We proposed a continuous-time model which allows to model the progression of myeloma disease in a cohort of patients. The interest of our method lies in the fact that we estimate not only the time to relapse, but also the time to entry into remission and the censoring indicator, enabling us to understand and estimate the distribution of time spent in remission. This is a different approach from that generally adopted by survival analysis or online-based prediction methods. These reconstructed durations can then be used by other models for further analysis, especially for non-parametric estimation and methods using covariates.  One limitation of our method is that it does not apply online. It requires the entire trajectory to have already been observed, and therefore cannot be plugged into a decision protocol as such. Our method is illustrated on a very simple type of trajectory, but it can easily be extended to more complex dynamics. In particular, it is not limited to exponential flow trajectories, nor to dimension 1, and can be adapted to detect more than two jumps.

\section{Data availability and code}
The data presented in this article and the code used to reproduce the results and the figures are available at \href{https://github.com/AmelieVernay/pdmp_relapse_time}{https://github.com/AmelieVernay/pdmp\_relapse\_time}.

\section{Fundings}

We acknowledge the support of European Union's Horizon 2020 research and innovation program (\href{https://marie-sklodowska-curie-actions.ec.europa.eu}{https://marie-sklodowska-curie-actions.ec.europa.eu}, Marie Sklodowska-Curie grant agreement No 890462, to Alice Cleynen), and of the French National Research Agency (ANR), under grant \href{https://anr.fr/Projet-ANR-20-CHIA-0001}{ANR-20-CHIA-0001} (https://anr.fr/, project CAMELOT, to Amélie Vernay). We also acknowledge the ANR under grant \href{https://anr.fr/Projet-ANR-21-CE40-0005}{ANR-21-CE40-0005} (https://anr.fr/, project HSMM-INCA).

\bibliographystyle{plainnat}
\bibliography{references}

@article{davis1984piecewise,
    author={Davis, MHA.},
    title={Piecewise-deterministic {M}arkov processes: a general class of nondiffusion stochastic models},
    note={With discussion},
    journal={J. Roy. Statist. Soc. Ser. B},
    fjournal={Journal of the Royal Statistical Society. Series B. Methodological},
    volume={46},
    year={1984},
    number={3},
    pages={353--388}
}

@article{blagoev2014therapies,
  title={Therapies with diverse mechanisms of action kill cells by a similar exponential process in advanced cancers},
  author={Blagoev, Krastan B and Wilkerson, Julia and Stein, Wilfred D and Yang, James and Bates, Susan E and Fojo, Tito},
  journal={Cancer research},
  volume={74},
  number={17},
  pages={4653--4662},
  year={2014},
  publisher={American Association for Cancer Research}
}

@article{sullivan1972kinetics,
  title={Kinetics of tumor growth and regression in IgG multiple myeloma},
  author={Sullivan, Peter W and Salmon, Sydney E and others},
  journal={The Journal of clinical investigation},
  volume={51},
  number={7},
  pages={1697--1708},
  year={1972},
  publisher={American Society for Clinical Investigation}
}

@article{attal2017lenalidomide,
    title={Lenalidomide, bortezomib, and dexamethasone with transplantation for myeloma},
    author={Attal, Michel and Lauwers-Cances, Valerie and Hulin, Cyrille and Leleu, Xavier and Caillot, Denis and Escoffre, Martine and Arnulf, Bertrand and Macro, Margaret and Belhadj, Karim and Garderet, Laurent and others},
    journal={New England Journal of Medicine},
    volume={376},
    number={14},
    pages={1311--1320},
    year={2017},
    publisher={Mass Medical Soc}
}

@article{truong2020review,
    title = {Selective review of offline change point detection methods},
    journal = {Signal Processing},
    volume = {167},
    pages = {107299},
    year = {2020},
    issn = {0165-1684},
    doi = {https://doi.org/10.1016/j.sigpro.2019.107299},
    url = {https://www.sciencedirect.com/science/article/pii/S0165168419303494},
    author = {Charles Truong and Laurent Oudre and Nicolas Vayatis}
}

@book{azais_statistical_2018,
	edition = {1},
	title = {Statistical {Inference} for {Piecewise}‐deterministic {Markov} {Processes}},
	isbn = {978-1-78630-302-8 978-1-119-50733-8},
	url = {https://onlinelibrary.wiley.com/doi/book/10.1002/9781119507338},
	language = {en},
	urldate = {2024-03-15},
	publisher = {Wiley},
	editor = {Azaïs, Romain and Bouguet, Florian},
	year = {2018},
	doi = {10.1002/9781119507338},
}

@article{krell_statistical_2016,
	title = {Statistical estimation of jump rates for a piecewise deterministic {Markov} processes with deterministic increasing motion and jump mechanism},
	volume = {20},
	issn = {1292-8100, 1262-3318},
	url = {http://www.esaim-ps.org/10.1051/ps/2016013},
	doi = {10.1051/ps/2016013},
	urldate = {2024-03-15},
	journal = {ESAIM: Probability and Statistics},
	author = {Krell, Nathalie},
	year = {2016},
	pages = {196--216},
}

@article{krell_nonparametric_2021,
	title = {Nonparametric estimation of jump rates for a specific class of piecewise deterministic {Markov} processes},
	volume = {27},
	issn = {1350-7265},
	url = {https://projecteuclid.org/journals/bernoulli/volume-27/issue-4/Nonparametric-estimation-of-jump-rates-for-a-specific-class-of/10.3150/20-BEJ1312.full},
	doi = {10.3150/20-BEJ1312},
	number = {4},
	urldate = {2024-12-30},
	journal = {Bernoulli},
	author = {Krell, Nathalie and Schmisser, Émeline},
	year = {2021},
}

@article{azais_nonparametric_2014,
	title = {Non‐{Parametric} {Estimation} of the {Conditional} {Distribution} of the {Interjumping} {Times} for {Piecewise}‐{Deterministic} {Markov} {Processes}},
	volume = {41},
	issn = {0303-6898, 1467-9469},
	url = {https://onlinelibrary.wiley.com/doi/10.1111/sjos.12076},
	doi = {10.1111/sjos.12076},
	language = {en},
	number = {4},
	urldate = {2024-03-08},
	journal = {Scandinavian Journal of Statistics},
	author = {Azaïs, Romain and Dufour, François and Gégout‐Petit, Anne},
	year = {2014},
	pages = {950--969},
}

@article{fonteijn_eventbased_2012,
    title = {An event-based model for disease progression and its application in familial Alzheimer's disease and Huntington's disease},
    journal = {NeuroImage},
    volume = {60},
    number = {3},
    pages = {1880-1889},
    year = {2012},
    issn = {1053-8119},
    doi = {https://doi.org/10.1016/j.neuroimage.2012.01.062},
    url = {https://www.sciencedirect.com/science/article/pii/S1053811912000791},
    author = {Hubert M. Fonteijn and Marc Modat and Matthew J. Clarkson and Josephine Barnes and Manja Lehmann and Nicola Z. Hobbs and Rachael I. Scahill and Sarah J. Tabrizi and Sebastien Ourselin and Nick C. Fox and Daniel C. Alexander}
}

@InProceedings{severson_personalized_2020,
  title = 	 {Personalized Input-Output Hidden Markov Models for Disease Progression Modeling},
  author =       {Severson, Kristen A. and Chahine, Lana M. and Smolensky, Luba and Ng, Kenney and Hu, Jianying and Ghosh, Soumya},
  booktitle = 	 {Proceedings of the 5th Machine Learning for Healthcare Conference},
  pages = 	 {309--330},
  year = 	 {2020},
  editor = 	 {Doshi-Velez, Finale and Fackler, Jim and Jung, Ken and Kale, David and Ranganath, Rajesh and Wallace, Byron and Wiens, Jenna},
  volume = 	 {126},
  series = 	 {Proceedings of Machine Learning Research},
  month = 	 {07--08 Aug},
  publisher =    {PMLR},
  url = 	 {https://proceedings.mlr.press/v126/severson20a.html}
}

@article{amoros_hmm_2019,
  author={Ruben Amoros and Ruth King and Hidenori Toyoda and Takashi Kumada and Philip J. Johnson and Thomas G. Bird},
  title={{A continuous-time hidden Markov model for cancer surveillance using serum biomarkers with application to hepatocellular carcinoma}},
  journal={METRON},
  year={2019},
  volume={77},
  number={2},
  pages={67-86},
  month={August},
  doi={10.1007/s40300-019-00151-},
  url={https://ideas.repec.org/a/spr/metron/v77y2019i2d10.1007_s40300-019-00151-8.html}
}

@article{drescher_screening_2013,
author = {Drescher, Charles W. and Shah, Chirag and Thorpe, Jason and O'Briant, Kathy and Anderson, Garnet L. and Berg, Christine D. and Urban, Nicole and McIntosh, Martin W. },
title = {Longitudinal Screening Algorithm That Incorporates Change Over Time in CA125 Levels Identifies Ovarian Cancer Earlier Than a Single-Threshold Rule},
journal = {Journal of Clinical Oncology},
volume = {31},
number = {3},
pages = {387-392},
year = {2013},
doi = {10.1200/JCO.2012.43.6691},

note ={PMID: 23248253},

URL = {https://ascopubs.org/doi/abs/10.1200/JCO.2012.43.6691},
eprint = {https://ascopubs.org/doi/pdf/10.1200/JCO.2012.43.6691},
}

@article{lorenzi_probabilistic_2019,
title = {Probabilistic disease progression modeling to characterize diagnostic uncertainty: Application to staging and prediction in Alzheimer's disease},
journal = {NeuroImage},
volume = {190},
pages = {56-68},
year = {2019},
note = {Mapping diseased brains},
issn = {1053-8119},
doi = {https://doi.org/10.1016/j.neuroimage.2017.08.059},
url = {https://www.sciencedirect.com/science/article/pii/S1053811917307061},
author = {Marco Lorenzi and Maurizio Filippone and Giovanni B. Frisoni and Daniel C. Alexander and Sebastien Ourselin}
}

@article{han_statistical_2020,
author = {Han, Yongli and Albert, Paul S. and Berg, Christine D. and Wentzensen, Nicolas and Katki, Hormuzd A. and Liu, Danping},
title = {Statistical approaches using longitudinal biomarkers for disease early detection: A comparison of methodologies},
journal = {Statistics in Medicine},
volume = {39},
number = {29},
pages = {4405-4420},
doi = {https://doi.org/10.1002/sim.8731},
url = {https://onlinelibrary.wiley.com/doi/abs/10.1002/sim.8731},
eprint = {https://onlinelibrary.wiley.com/doi/pdf/10.1002/sim.8731},
year = {2020}
}

@article{tang_biomarker_2017,
author = {Xiaoying Tang and Michael I. Miller and Laurent Younes},
title = {{Biomarker change-point estimation with right censoring in longitudinal studies}},
volume = {11},
journal = {The Annals of Applied Statistics},
number = {3},
publisher = {Institute of Mathematical Statistics},
pages = {1738 -- 1762},
year = {2017},
doi = {10.1214/17-AOAS1056},
URL = {https://doi.org/10.1214/17-AOAS1056}
}

@article{delft_modeling_2022,
title = {Modeling strategies to analyse longitudinal biomarker data: An illustration on predicting immunotherapy non-response in non-small cell lung cancer},
journal = {Heliyon},
volume = {8},
number = {10},
pages = {e10932},
year = {2022},
issn = {2405-8440},
doi = {https://doi.org/10.1016/j.heliyon.2022.e10932},
url = {https://www.sciencedirect.com/science/article/pii/S2405844022022204},
author = {Frederik A. {van Delft} and Milou Schuurbiers and Mirte Muller and Sjaak A. Burgers and Huub H. {van Rossum} and Maarten J. IJzerman and Hendrik Koffijberg and Michel M. {van den Heuvel}}
}

@article{bartolomeo_progression_2011,
author={Bartolomeo, Nicola and Trerotoli, Paolo and Serio, Gabriella},
title={Progression of liver cirrhosis to HCC: an application of hidden Markov model},
journal={BMC Medical Research Methodology},
year={2011},
month={Apr},
day={04},
volume={11},
number={1},
pages={38},
issn={1471-2288},
doi={10.1186/1471-2288-11-38},
url={https://doi.org/10.1186/1471-2288-11-38}
}

@misc{hmmlearn,
  author = {Weiss, Ron and Du, Shiqiao and Grobler, Jaques and Cournapeau, David and Pedregosa, Fabian and Varoquaux, Gael and Mueller, Andreas and Thirion, Bertrand and Nouri, Daniel and Louppe, Gilles and Vanderplas, Jake and Benediktsson, John and Buitinck, Lars and Korobov, Mikhail and McGibbon, Robert and Lattarini, Stefano and Niculae, Vlad and csytracy and Gramfort, Alexandre and Lebedev, Sergei and Huppenkothen, Daniela and Farrow, Christopher and Yanenko, Alexandr and Lee, Antony and Danielson, Matthew and Rockhill, Alex},
  title = {hmmlearn},
  url = {https://github.com/hmmlearn/hmmlearn},
  version = {0.0.3},
  year = {2024}
}

@article{cloez2017probabilistic,
  title={Probabilistic and piecewise deterministic models in biology},
  author={Cloez, Bertrand and Dessalles, Renaud and Genadot, Alexandre and Malrieu, Florent and Marguet, Aline and Yvinec, Romain},
  journal={ESAIM: Proceedings and Surveys},
  volume={60},
  pages={225--245},
  year={2017},
  publisher={EDP Sciences}
}

@misc{azais2025asymptotic,
      title={Asymptotic Analysis and Practical Evaluation of Jump Rate Estimators in Piecewise-Deterministic Markov Processes}, 
      author={Azaïs, Romain  and Denis, Solune},
      year={2025},
      eprint={2502.14621},
      archivePrefix={arXiv},
      primaryClass={stat.ME},
      url={https://arxiv.org/abs/2502.14621}, 
}

@article{pilgrim2021piecewise,
    doi = {10.21105/joss.03859},
    url = {https://doi.org/10.21105/joss.03859},
    year = {2021},
    publisher = {The Open Journal},
    volume = {6},
    number = {68},
    pages = {3859},
    author = {Pilgrim, Charlie},
    title = {piecewise-regression (aka segmented regression) in Python},
    journal = {Journal of Open Source Software}
}

@book{elliott1997hmm,
    author = {Elliott,Robert James and Aggoun,Lakhdar and Moore,John B.},
    booktitle = {Hidden Markov models : estimation and control},
    edition = {Corr. 2e éd.},
    isbn = {978-0-387-94364-0},
    publisher = {Springer-Verlag},
    series = {Applications of mathematics},
    title = {Hidden Markov models : estimation and control / Robert J. Elliott, Lakhdar Aggoun, John B. Moore},
    year = {1995},
}

@article{aminikhanghahi2016survey,
  title={A survey of methods for time series change point detection},
  author={Samaneh Aminikhanghahi and Diane Joyce Cook},
  journal={Knowledge and Information Systems},
  year={2016},
  volume={51},
  pages={339 - 367},
  url={https://api.semanticscholar.org/CorpusID:15595198}
}

\end{document}